\documentclass[journal]{IEEEtran}
\makeatletter
\newcommand*{\rom}[1]{\expandafter\@slowromancap\romannumeral #1@}
\makeatother
\usepackage{amsmath, amssymb, bm, cite, epsfig, psfrag, ma thtools}
\usepackage{graphicx}
\usepackage[margin=0.468in]{geometry}

\usepackage{dblfloatfix}
\usepackage{array}
\usepackage{lipsum}
\usepackage{array}
\newcolumntype{P}[1]{>{\centering\hspace{0pt}}p{#1}}
\newcolumntype{M}[1]{>{\centering\hspace{0pt}}m{#1}}
\newcolumntype{L}{>{\centering\arraybackslash}m{3cm}}
\usepackage[font=small]{caption}
\usepackage{subcaption}
\usepackage{epstopdf}
\usepackage{longtable}
\usepackage{supertabular,booktabs}
\usepackage{capt-of}
\usepackage{bbm}
\usepackage{multirow}
\usepackage[usenames,dvipsnames]{xcolor}

\renewcommand{\arraystretch}{1.5}
\usepackage{etoolbox}
\usepackage{pbox}
\usepackage{fixltx2e}
\usepackage[hyphens]{url}
\usepackage{colortbl}
\usepackage{fancyhdr}

\def\dB{\mathrm{dB}}
\DeclareMathOperator{\sinc}{sinc}
\usepackage{tabu}
\newcommand{\squeezeup}{\vspace{-2.5mm}}
\usepackage{tikz}
\usetikzlibrary{calc}

\usepackage{bm}
\newtoggle{conference}
\togglefalse{conference} 
\graphicspath{{figures/}}

\fancypagestyle{plain}{
	\fancyhf{}
	\fancyhead[C]{Conference on \LaTeX}             
	\fancyfoot[L]{This is a notice}        

}

\newcommand*{\balancecolsandclearpage}{%
	\close@column@grid
	\cleardoublepage
	\twocolumngrid
}

\begin{document}
\bibliographystyle{IEEEtran}
\title{\huge Small-Scale, Local Area, and Transitional Millimeter Wave Propagation for 5G Communications}
\author{\IEEEauthorblockN{Theodore S. Rappaport, George R. MacCartney Jr., Shu Sun, Hangsong Yan, and Sijia Deng}\vspace{-0.7cm}
\thanks{This material is based upon work supported by the NYU WIRELESS Industrial Affiliates program, NSF research grants 1320472, 1302336, and 1555332, and funding from Nokia. The authors acknowledge Amitava Ghosh of Nokia for his support of this work. The authors thank Yunchou Xing, Jeton Koka, Ruichen Wang, and Dian Yu for their help in conducting the measurements. The authors also thank Prof. Henry Bertoni for his insights and advisement in diffraction theory.}
\thanks{T. S. Rappaport (email: tsr@nyu.edu), G. R. MacCartney, Jr., S. Sun, H. Yan, and S. Deng are with NYU WIRELESS Research Center, NYU Tandon School of Engineering, 9th Floor, 2 MetroTech Center,  Brooklyn, NY 11201.}}

\maketitle
\begin{tikzpicture}[remember picture, overlay]
\node[font=\small] at ($(current page.north) + (0,-0.2in)$) {T. S. Rappaport, G. R. MacCartney, Jr., S. Sun, H. Yan, S. Deng, ``Small-Scale, Local Area, and Transitional Millimeter Wave Propagation};
\node[font=\small] at ($(current page.north) + (0,-0.35in)$)  {for 5G Communications," in \textit{IEEE Transactions on Antennas and Propagation, Special Issue on 5G}, Nov. 2017. };
\end{tikzpicture}
\thispagestyle{empty}
\begin{abstract}
This paper studies radio propagation mechanisms that impact handoffs, air interface design, beam steering, and MIMO for 5G mobile communication systems. Knife edge diffraction (KED) and a creeping wave linear model are shown to predict diffraction loss around typical building objects from 10 to 26 GHz, and human blockage measurements at 73 GHz are shown to fit a double knife-edge diffraction (DKED) model which incorporates antenna gains. Small-scale spatial fading of millimeter wave received signal voltage amplitude is generally Ricean-distributed for both omnidirectional and directional receive antenna patterns under both line-of-sight (LOS) and non-line-of-sight (NLOS) conditions in most cases, although the log-normal distribution fits measured data better for the omnidirectional receive antenna pattern in the NLOS environment. Small-scale spatial autocorrelations of received voltage amplitudes are shown to fit sinusoidal exponential and exponential functions for LOS and NLOS environments, respectively, with small decorrelation distances of 0.27 cm to 13.6 cm (smaller than the size of a handset) that are favorable for spatial multiplexing. Local area measurements using cluster and route scenarios show how the received signal changes as the mobile moves and transitions from LOS to NLOS locations, with reasonably stationary signal levels within clusters. Wideband mmWave power levels are shown to fade from 0.4 dB/ms to 40 dB/s, depending on travel speed and surroundings. 
\end{abstract}

\section{Introduction}~\label{sec:intro}
Driven mainly by the pervasive usage of smartphones and the emergence of the Internet of Things (IoT), future 5G mobile networks will become as pervasive as electrical wiring~\cite{Rap91d} and will offer unprecedented data rates and ultra-low latency~\cite{Boccardi14a,Ghosh14a,Rap16a}. For the first time in the history of radio, millimeter-wave (mmWave) frequencies will be used extensively for mobile and fixed access, thus requiring accurate propagation models that predict how the channel varies as people move about. Remarkable progress has been made in modeling large-scale propagation path loss at mmWave frequencies~\cite{Rap13a,Rap15b,Rap16a, Rap13b,Mac15b,Sun16b,Thomas16a,Koymen15a,Yoon15a,Maltsev10a,Mac17b}, and it is well understood that for an assumption of unity gain antennas across all frequencies, Friis equation predicts that path loss is greater at mmWave compared to today's UHF/microwave cellular systems~\cite{Friis46a, Rap13a, Rap15b, Mac15b,Sun16b,Thomas16a,Rap16a}. Also, rain and atmospheric attenuation are well understood, and reflection and scattering are more dominant than diffraction at mmWave bands~\cite{Rap16a,Rap15a,Deng16a,Rumney16b,Solomitckii16a}.

Broadband statistical spatial channel models (SSCMs) and simulators that faithfully predict the statistics of signal strength, and the number and direction of arrival and departure of multipath components, have been developed by a consortium of companies and universities~\cite{3GPP.38.900} and from measurements in New York City\textcolor{black}{\cite{Sun17b}}. These models are being used to develop air-interfaces for 5G systems~\cite{Samimi16a}\textcolor{black}{\cite{Sun17c}}. Elsewhere in this issue,\textcolor{black}{\cite{Rap16a}} summarizes standard activities for large-scale mmWave channel modeling.

Little is known, however, about the small-scale behavior of wideband mmWave signals as a mobile user moves about a local area. Such information is vital for the design of handoff mechanisms and beam steering needed to rescue the communication link from deep fades. In this paper, propagation measurements investigate diffraction, human blocking effects, small-scale spatial fading and autocorrelation, local area channel transitions, and stationarity of signal power in local area clusters at frequencies ranging from 10 to 73 GHz. Diffraction measurements for indoor and outdoor materials at 10, 20, and 26 GHz are presented in Section II, and two diffraction models, i.e., the knife edge diffraction (KED) model and a creeping wave linear model, are used to fit the measured results. We predict the rapid signal decay as a mobile moves around a diffracting corner. In Section III, measurements at 73 GHz are presented and a double knife-edge diffraction (DKED) antenna gain model that uses directional antenna patterns is shown to describe minimum and maximum fade depths caused by human blockage. Small-scale fading and correlation studies at 73 GHz are presented in Section IV, where small-scale fading distributions and spatial autocorrelations of received voltage amplitudes in LOS and NLOS environments with omnidirectional and directional antennas are provided and analyzed. In Section V, \textit{route} and \textit{cluster} scenarios are used to study local area channel transitions and stationarity, where analysis for channel transition from a NLOS to a LOS region and local area path loss variations are provided. Conclusions are given in Section VI. \textcolor{black}{Channel models given here may be implemented for small-scale propagation modeling and real-time site-specific mobile channel prediction and network control~\cite{Wang05b}.}
                        
\section{Diffraction Measurements and Models}\label{sec:diffraction}
\subsection{Introduction of Diffraction Measurements and Models}
Accurate characterization of diffraction at cmWave and mmWave frequencies is important for understanding the rate of change of signal strength for mobile communications since future 5G mmWave systems will have to rely less on diffraction as a dominant propagation mechanism~\cite{Rap15b, Mac15b}. Published indoor and outdoor diffraction measurements show that diffraction has little contribution to the received signal power using various materials and geometries (edges, wedges, and circular cylinders) at 60 GHz and 300 GHz~\cite{Maltsev10c,Lu13a,Jacob12a}. It was shown that the KED model agreed well with diffraction measurements for cuboids at 300 GHz~\cite{Kleine-Ostmann12a}, vegetation obstacles at 2.4, 5, 28, and 60 GHz~\cite{Corre16a}, and for human blocking at 60 GHz~\cite{Maltsev10a}. Apart from the KED model, uniform theory of diffraction (UTD) models are also used. An overview of the Geometrical Theory of Diffraction (GTD) and the UTD are provided in~\cite{Pathak13a}, as well as their utility to solve practical problems. Besides the KED and UTD models, Mavridis \textit{et al.} presented a creeping wave linear model~\cite{Mavridis14a} to estimate the diffraction loss by a perfectly conducting or lossy circular cylinder for both transverse-magnetic (TM) and transverse-electric (TE) polarizations at 60 GHz. In the following sub-sections, we describe diffraction measurements conducted in 2015 around the engineering campus of New York University~\cite{Deng16a}, where the frequency dependency of diffraction at 10, 20, and 26 GHz in realistic indoor and outdoor scenarios was investigated to yield simple yet accurate diffraction models for wireless planning.
      
\subsection{Diffraction Measurement System}
Diffraction measurements were performed by transmitting a continuous wave (CW) signal generated by an Agilent E8257D PSG analog signal generator through a pyramidal horn antenna at the transmitter (TX). An identical horn antenna was used at the receiver (RX) to receive signal energy around a corner test material (e.g. a stone pillar). The RX antenna was fed to an E4407B ESA-E spectrum analyzer that measured received power which was subsequently recorded on a laptop with LabVIEW software. During the measurements, the TX antenna was set one meter from the knife edge, sufficiently in the far field, and was fixed to a tripod and aimed at the knife edge, whereas the RX antenna was set 2 meters from the knife edge (also in the far field) and was fixed on a rotatable gimbal attached to a translatable linear track that was made to from an approximate arc around the knife edge (see Fig. \ref{fig:Corner}). Diffraction loss was measured at 10, 20, and 26 GHz using identical pairs of antennas at the TX and RX to measure each frequency, separately. For each frequency, Table \ref{tbl:1} lists the flange type, antenna gain, and half-power beamwidth (HPBW) of the antenna pairs used. The TX and RX were stationed at a wide range of angles, both in the lit and shadowed region, and the horns always had their boresights focused on the corner knife-edge of the indoor and outdoor materials that were studied. More details regarding the measurement system are given in~\cite{Deng16a,Deng16b}.

\begin{table}[b]
	\renewcommand{\arraystretch}{1.4}
	\begin{center}
		\caption{Antenna parameters used diffraction measurements.}~\label{tbl:1}
		\fontsize{7}{7}\selectfont
		\begin{tabular}{|>{\centering\arraybackslash}m{1.3cm}|>{\centering\arraybackslash}m{1.1cm}|>{\centering\arraybackslash}m{1.4cm}|>{\centering\arraybackslash}m{1.25cm}|>{\centering\arraybackslash}m{1.5cm}|>{\centering\arraybackslash}m{0.6cm}|}\hline
			\textbf{Measured Frequency}	& \textbf{Flange Type}	& \textbf{Antenna Gain}	& \textbf{HPBW (Az./El.)}	& \textbf{Far Field Distance}	\\ \hline
			10 GHz			& WR-75 		& 20 dBi			& $17^\circ$/$17^\circ$					& 0.47 m		\\ \hline
			20 GHz 			& WR-51 		& 20 dBi 			& $17^\circ$/$17^\circ$				    & 0.46 m		\\ \hline
			26 GHz 			& WR-28 		& 24.5 dBi 			& $10.9^\circ$/$8.6^\circ$					& 0.83 m		\\ \hline
		\end{tabular}
	\end{center}
\end{table}

\subsection{Diffraction Measurement Description}
\begin{figure}[b!]
	\squeezeup
	\begin{center}
		\includegraphics [width=0.44\textwidth]{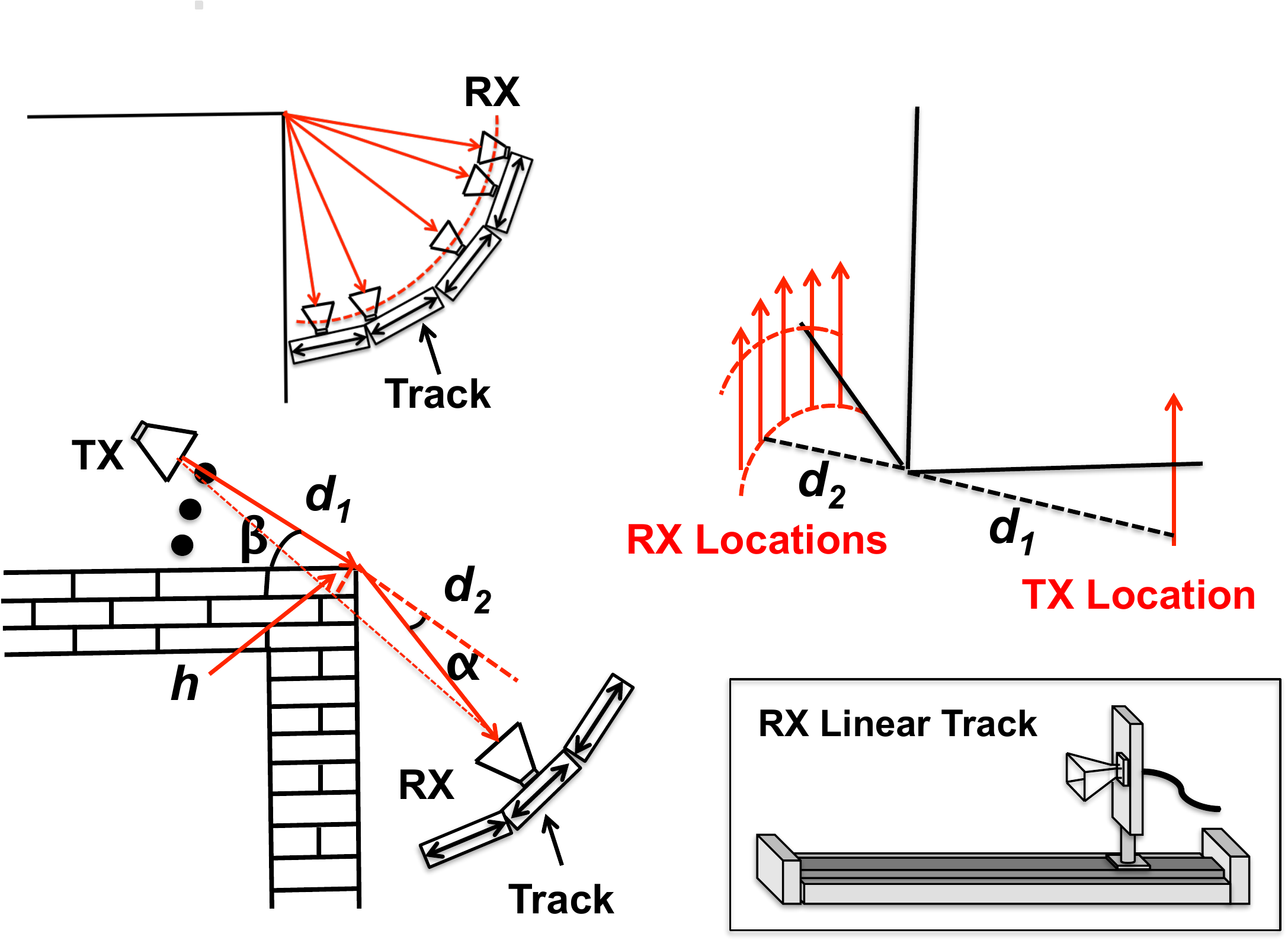}
		\caption{Top view of the corner diffraction geometry\cite{Deng16a}.}\label{fig:Corner}
	\end{center}
\end{figure}

The indoor measurements were performed at $90^\circ$ (right-angle) wall corners made of drywall, wood, and semi-transparent plastic board with 2 cm thickness. Outdoor measurements studied one rounded stone pillar corner and one right-angle marble building corner. During the measurements, the TX and RX were placed on either side of the corner (knife edge) of the test material. A diagram of the corner diffraction geometry is shown in Fig. \ref{fig:Corner} where $d_1$ is the distance between the TX and the corner knife-edge, and $d_2$ is the distance between the corner knife-edge and the RX. Both $d_1$ (1 m) and $d_2$ (2 m) remained constant throughout the diffraction measurement campaign. The $\beta$ and $\alpha$ values are the incident and diffraction angles, respectively, where two (outdoor) or three (indoor) fixed values between $10^\circ$ and $39^\circ$ were chosen for $\beta$. The RX antenna was mounted on a motorized linear track (see Fig. \ref{fig:Corner}) that translated in step increments of 0.875 cm, which corresponds to approximately a $0.5^\circ$ increment in diffraction angle ($\alpha$) for each step increment. At each step increment, the RX antenna was adjusted to point perfectly towards the knife-edge corner. The length of the track was $35.5$ cm and was used to measure a $20^\circ$ swath of diffraction angles over the entire length of the track. At each measurement location, five consecutive linear tracks (see Fig. \ref{fig:Corner}) were used to provide a $100^\circ$ diffraction angle arc around the corner which covered a broad range of the shadow region where the TX antenna is shadowed by the corner object with respect to RX antenna. Additionally, a smaller range of diffraction angles was measured in the lit region where the TX and RX antennas were in view of each other but were not pointed at each other since they were always aimed at the corner. 

At each location, prior to the diffraction measurements, a free space calibration in an open area with both antennas pointed at each other on boresight was conducted with a 3 m ($d_1 + d_2 = 3$ m) transmitter-receiver (T-R) separation distance to provide a free space power reference for each frequency. The diffraction loss was then obtained by calculating the difference between the measured received signal power at each step increment of the RX antenna during the diffraction measurements and the free space calibration received power (the TX and RX antenna gains were deducted from all power measurements).
                 
\subsection{Theoretical Diffraction Models}
\subsubsection{KED Model}
The KED model is suitable for applications with sharp knife edges and has a simple form yet high prediction accuracy ~\cite{Russell93a}. In general, diffraction loss over complex and irregular obstructions can be difficult to calculate, but typical obstructions for 5G wireless will involve common building partitions which are generally simple in nature and with dimensions that appear infinite at such small wavelengths, such as a wall or building corner, thus justifying simple diffraction models. 

The diffraction loss (as compared to free space) is obtained by calculating the electric field strength $E_d$ [V/m] at the RX based on the specific Fresnel diffraction parameter $\nu$~\cite{Rap02a}. The ratio of $E_d$ and the free space field strength $E_0$ can be computed by summing all the secondary \textcolor{black}{Huygens'} sources in the knife edge plane and is given by~\cite{Deng16a,Rap02a}:
\begin{equation}
\footnotesize
\label{eq:8}
\frac{E_d}{E_0} = F(\nu) =  \frac{1+j}{2}  \int_{\textcolor{black}{\nu}}^{\infty} e^{-j (\pi/2) t^2}dt
\end{equation}
where $F(\nu)$ is the complex Fresnel integral and $\nu$ is the Fresnel diffraction parameter is defined as~\cite{Rap02a}:
\begin{equation}
\label{eq:9}
\footnotesize
\nu=h\sqrt{\frac{2(d_1+d_2)}{\lambda d_1 d_2}}=\alpha \sqrt{\frac{2 d_1 d_2}{\lambda(d_1+d_2)}}
\end{equation}
where $\lambda$ is the wavelength, $\alpha$ is the diffraction angle, $h$ is the effective height (or width) of the obstructing screen with an infinite width (or height) placed between the TX and RX at the distances $d_1$ and $d_2$, respectively, under the conditions that $d_1$, $d_2 \textcolor{black}{\gg} h$, and  $d_1$, $d_2 \textcolor{black}{\gg} \lambda$. These conditions were met for 10, 20 and 26 GHz measurements in both the indoor and outdoor environments, as shown in Fig.~\ref{fig:Corner}.

Based on (\ref{eq:8}) and (\ref{eq:9}), the diffraction power gain $G(\nu)$ in dB produced in a knife edge by the KED model is expressed as~\cite{Deng16a,Rap02a}:
\begin{equation}
\label{eq:10}
\footnotesize
G(\nu) [\dB] = - P(\nu) [\dB] = 20\log_{10}|F(\nu)|
\end{equation}
where $P(\nu)$ is the power loss of the diffracted signal for the value of $\nu$, compared to free space case for the same distance.

\subsubsection{Convex Surfaces based Diffraction Model}       
Although the KED model has broad applications for various geometries, it requires the diffraction corner to be in the shape of a sharp knife edge and does not account for the radius of curvature of an obstacle. When a diffraction corner is rounded in shape, such as a stone pillar corner (it resembles a circular cylinder), a creeping wave linear model can better predict diffraction loss~\cite{Piazzi98a, Mavridis14a}. A creeping ray field at the RX antenna behind the circular object for an incident plane wave is given by~\cite{Piazzi98a}:
\begin{equation} \label{eq:11}
\footnotesize
E(\alpha,d_2,k) \sim E_i e ^{-jk\alpha R_h} \frac{e^{-jkd_2}}{\sqrt{kd_2}}\sum_{p=1}^{\infty}{D_p{R_h}}\cdot \exp(-\psi_{p}\alpha)
\end{equation}
where $E_i$ is the incident field from the TX that impinges upon the obstruction, $R_h$ is the radius of cylinder for the diffraction corner, $k$ is the wave number of the carrier frequency, $\alpha$ is the diffraction angle, $d_2$ is the distance between the launch point at the rounded corner edge and the RX, $D_p$ is the excitation coefficient and $\psi_p$ is the attenuation constant. Due to the \textcolor{black}{computational complexity} of (\ref{eq:11}), a reasonable approximation for $E$ on a flat surface can be obtained by considering only the $p = 1$ term in (\ref{eq:11}), which is given by~\cite{Deng16a,Deng16b,Mavridis14a,Piazzi98a,Tervo14a}:
\begin{equation} \label{eq:12}
\footnotesize
E \sim E_i D_p{R_h} \cdot \exp(-\psi_{p}\alpha)
\end{equation}
The expression for the diffraction power loss in dB based on the creeping wave linear model is given by~\cite{Deng16a,Deng16b}:
\begin{equation}\label{eq:13}
\footnotesize
\begin{split}
G(\alpha) [\dB] = - P(\alpha) [\dB]= 20\log_{10}E
\end{split}
\end{equation}
In order to facilitate the computation of $G(\alpha)$, \textcolor{black}{a simple linear model}~\eqref{eq:14} based on minimum \textcolor{black}{mean squared} error (MMSE) estimation between the model and measured data was proposed in~\cite{Tervo14a} to estimate the diffraction loss caused by a curved surface at a single frequency, based on the creeping wave linear model~\cite{Deng16a,Deng16b}:
\begin{equation}
\footnotesize
\label{eq:14}
P(\alpha) = n \cdot \alpha + c
\end{equation}
where $n$ is the linear slope of diffraction loss calculated by MMSE for each specific frequency and object radius, and $c$ is the anchor point set to 6.03 dB, the diffraction loss estimated by KED with diffraction angles $\alpha=\beta=0^\circ$.      
\begin{figure}[t!]
	\centering
	\includegraphics[width=2.95in]{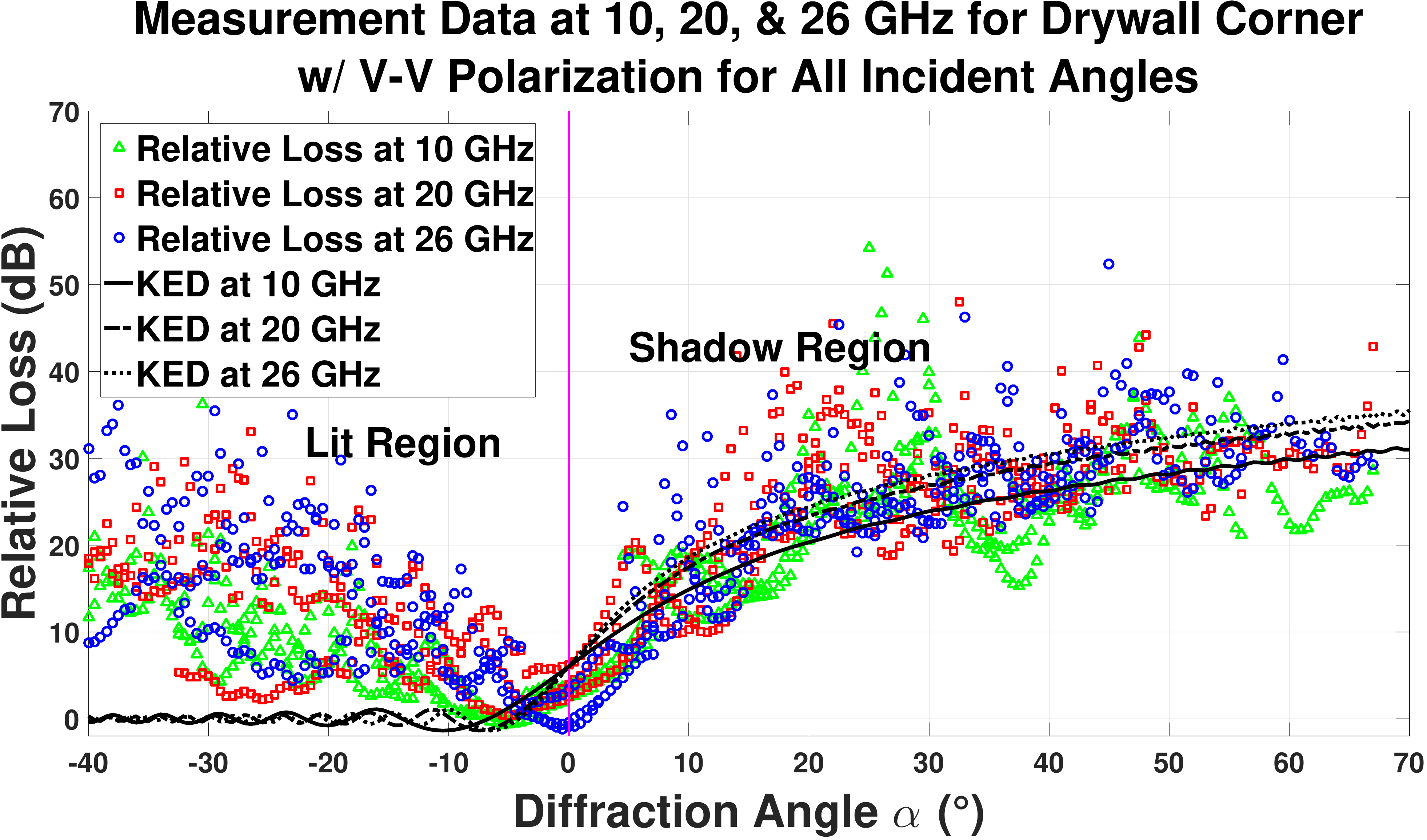}\vspace{0.2cm}
	\includegraphics[width=2.95in]{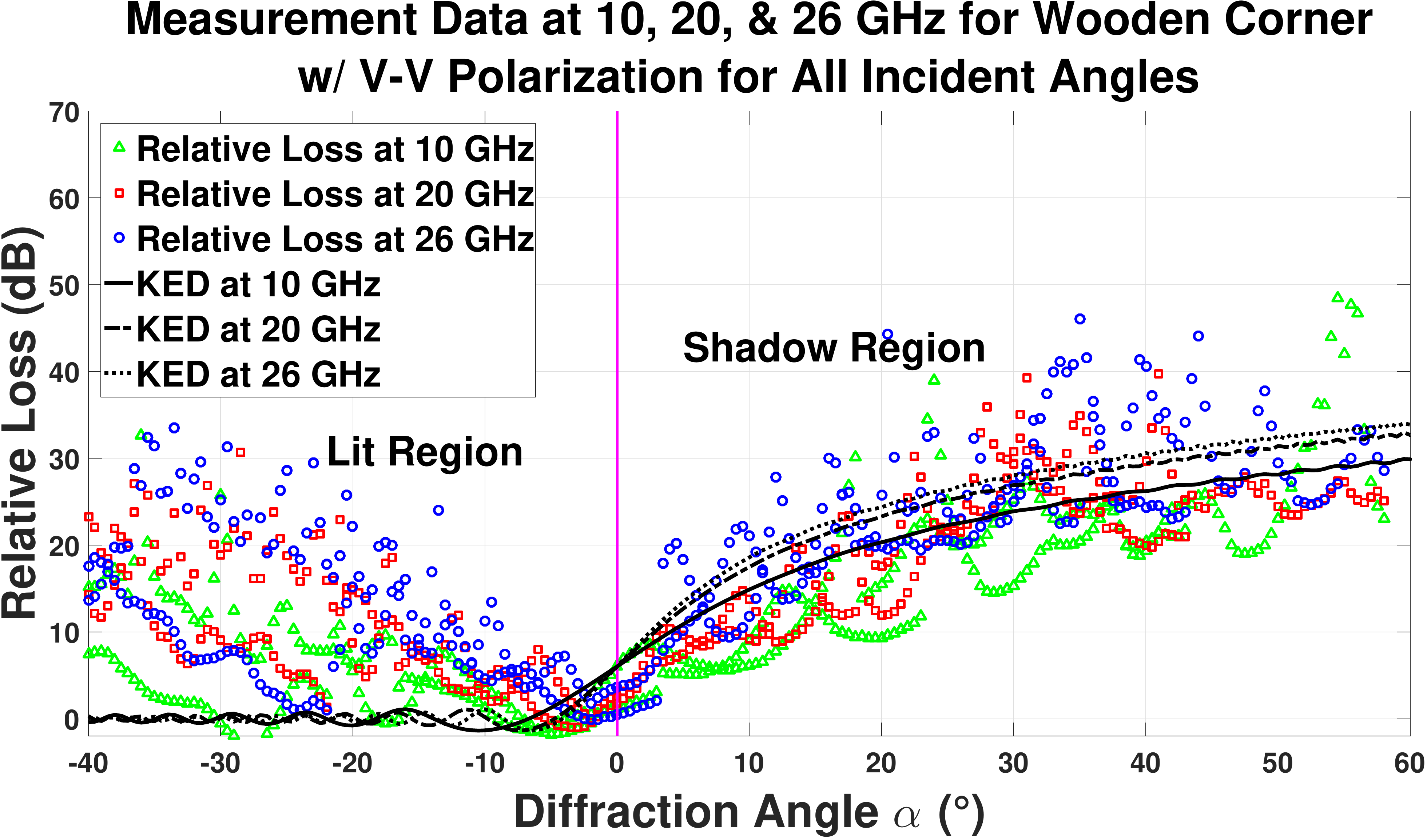}\vspace{0.2cm}
	\includegraphics[width=2.95in]{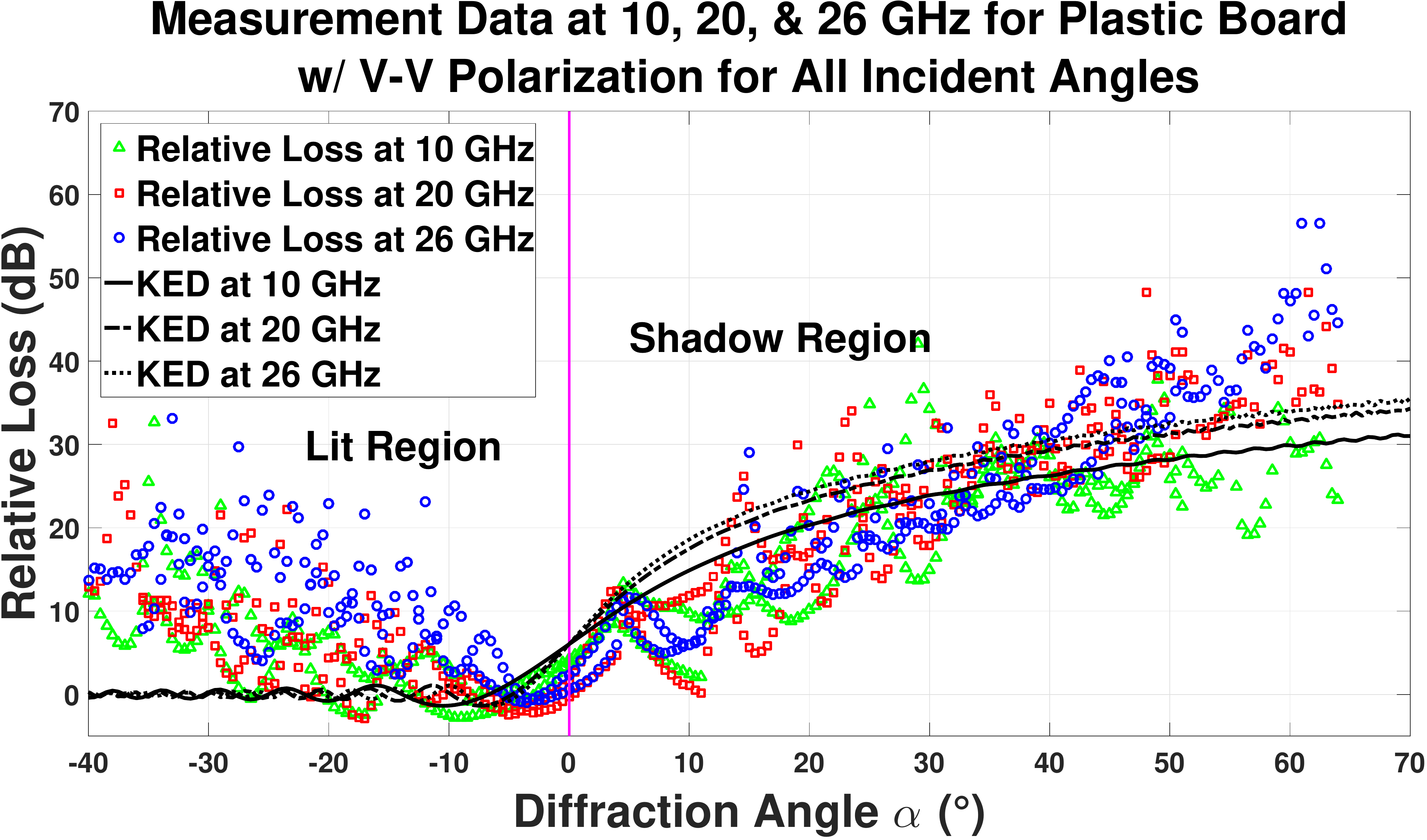}
	\caption{Diffraction measurements for the indoor drywall corner (top), wooden corner (middle), and plastic board (bottom) compared to the KED model at 10, 20 and 26 GHz~\cite{Deng16a}.}
	\label{fig:Indoor}
\end{figure}
\begin{figure}[b!]
	\centering
	\includegraphics[width=2.9in]{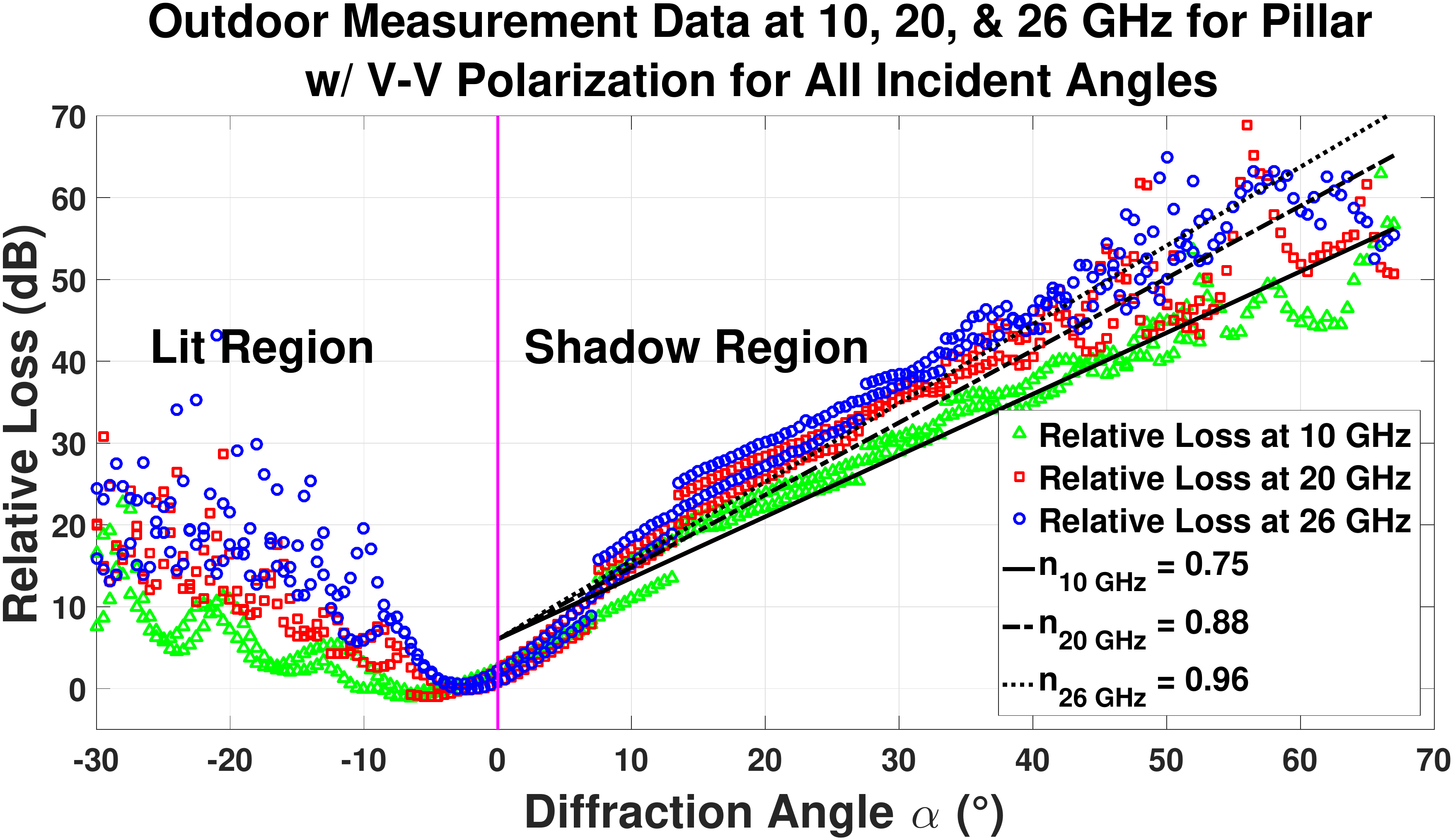}\vspace{0.2cm}
	\includegraphics[width=2.9in]{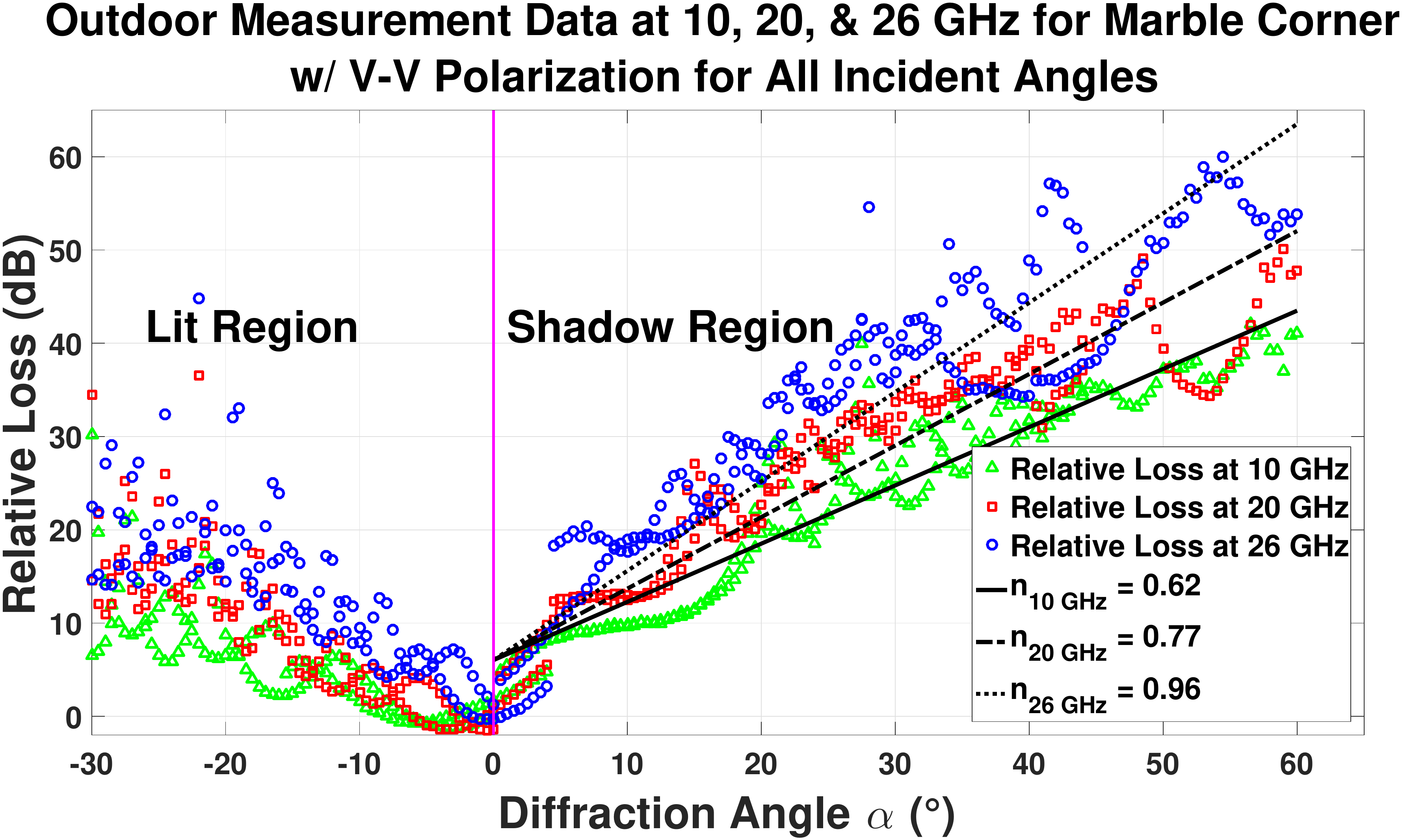}\vspace{0.2cm}
	\includegraphics[width=2.9in]{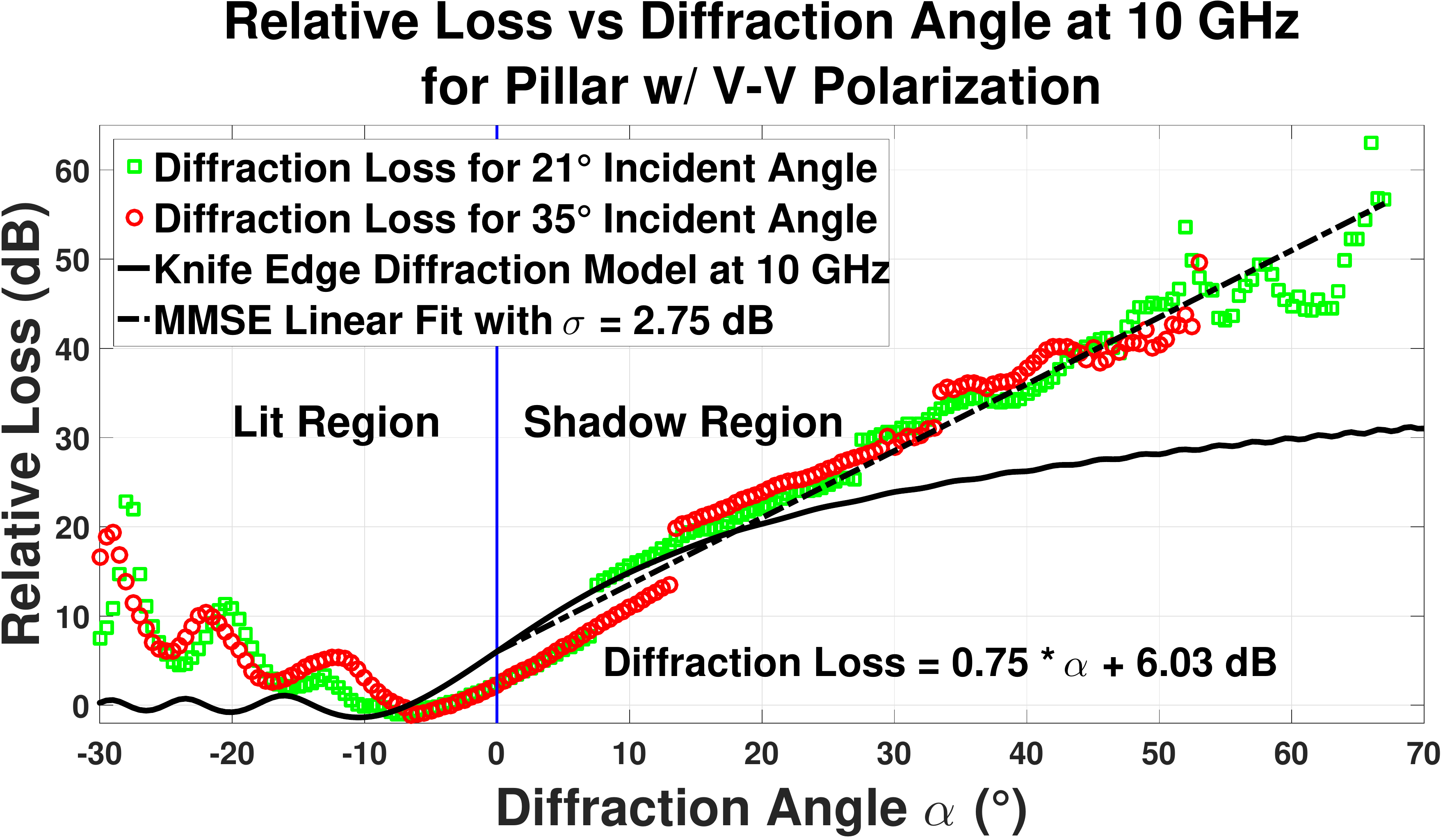}
	\caption{Measured diffraction loss for the outdoor stone pillar (top) and marble corner (middle) with creeping wave linear models at 10, 20, and 26 GHz, and stone pillar measurement results compared to the KED and the creeping wave linear models at 10 GHz (bottom)~\cite{Deng16a}.}
	\label{fig:Outdoor}
\end{figure}

\subsection{Indoor Diffraction Results and Analysis}
The indoor diffraction loss measurements for the drywall corner, wooden corner, and plastic board are plotted with the KED model (\ref{eq:10}) at 10, 20, and 26 GHz as a function of the diffraction angle in Fig. \ref{fig:Indoor}. \textcolor{black}{Three different TX incident angles were used to measure diffraction loss for each frequency. Since the measurements in~\cite{Deng16a} were conducted with the TX and RX along a constant radius ($d_1$ and $d_2$, respectively) from the corner of each test material, diffraction loss can be represented as a function of the diffraction angle $\alpha$ in~\eqref{eq:9} for each frequency (or wavelength $\lambda$), without the need for the TX incident angle.}. Fig. \ref{fig:Indoor} shows diffraction loss increases to approximately 30 dB as compared to free space as the RX antenna moves from the edge of the lit region (0$^\circ$) into the shadow region (20$^\circ$) for the drywall corner and wooden corner, respectively. This demonstrates the rapid signal degradation that occurs when diffraction is the primary propagation mechanism in mobile systems. Note the good fit between the drywall diffraction measurements and the simple KED model in the early and deep shadow regions for all three frequencies, while the KED model overestimates diffraction loss by 5-10 dB compared with the wooden corner diffraction measurements at 10, 20, and 26 GHz.

As for the plastic board material, the KED model overestimates the measured diffraction loss at small diffraction angles near the lit/shadow region boundary (from $0^\circ$ to $30^\circ$), and underestimates diffraction loss in portions of the deep shadow region (diffraction angles greater than $30^\circ$), most noticeably at 20 GHz and 26 GHz. Fig. \ref{fig:Indoor} indicates slightly less loss occurs at lower frequencies, implying frequency dependence, where divergence from pure diffraction theory  can be attributed to reflections and scattering in the indoor environment and potential transmissions through the test materials. In general, the observations match the KED model trend. Oscillation patterns of the measured data in the shadow region observed in Fig. \ref{fig:Indoor} indicate that the measured diffraction signal includes corner diffraction, penetration through the material, and partial scattering in the measurement environment. We note that the diffraction loss observed in the lit region is due to the measurement procedure where the TX and RX antennas were never aligned on boresight (except at 0$^\circ$) since they were constantly pointed directly at the knife-edge corner.  

\textcolor{black}{Penetration loss was measured for typical building materials in the same indoor environment at 73 GHz and showed that co-polarized penetration loss ranged from 0.8 dB/cm (lowest loss material -- drywall) to 9.9 dB/cm (highest loss material -- steel door), with standard deviation $\sigma$ about the average loss ranging from 0.3 dB/cm (lowest $\sigma$ -- drywall) to 2.3 dB/cm (highest $\sigma$ -- clear glass). Additional details can be found in Table II of~\cite{Ryan17a}.}
 
\subsection{Outdoor Diffraction Results and Analysis}
The outdoor marble corner and stone pillar measurement results are shown in Fig. \ref{fig:Outdoor} with the MMSE creeping wave linear models (\ref{eq:14}) at 10, 20, and 26 GHz.
It can be seen from Fig. \ref{fig:Outdoor} that (\ref{eq:14}) predicts a linearly increasing diffraction loss into the deeply shadowed region, as opposed to the leveling off seen in Fig. \ref{fig:Indoor} from (\ref{eq:10}).
Fig. \ref{fig:Outdoor} also shows the outdoor stone pillar measurement results at 10 GHz compared to the KED model and creeping wave linear model, where the creeping wave linear model provides a better fit to the measured relative diffraction loss than the KED model, in the shadow region.   
Diffraction loss for each frequency is plotted as a function of diffraction angle and includes the measured loss at two TX incident angles. 
The measured data matches well with the creeping wave linear model derived via MMSE for each frequency, in the shadow region. 
The creeping wave linear model slopes are 0.75, 0.88, and 0.96 for the stone pillar measurements and 0.62, 0.77, and 0.96 for the marble corner measurements at 10, 20, and 26 GHz, respectively.
Fig. \ref{fig:Outdoor} shows outdoor obstructions cause an even greater loss when a mobile moves into a deeply shadowed region, showing as much as 50 dB of loss when solely based on diffraction. 
Based on the increase in slope values with frequency, it is easily seen that diffraction loss increases with frequency in the outdoor environment. 
The simple creeping wave linear model (\ref{eq:14}) fits well with the measured data and has a much lower standard deviation compared with the KED model~\cite{Deng16a}, indicating a good overall match between the creeping wave linear model and measured data. 
Similar to the indoor environment diffraction loss measurements, the high diffraction loss in the lit region is caused by the TX and RX pointing off boresight towards the corner of the test material.

When comparing the slope values of the two outdoor materials, the rougher surface (stone) with a slightly rounded edge has a greater slope (greater attenuation) for identical frequencies as compared to the smoother surface (marble) straight edge. 
We note that using the MMSE method to derive the typical slope values instead of calculating the theoretical slope value provides system engineers a useful parameter while reducing the \textcolor{black}{computational complexity} of the diffraction model. The simple slope values are useful for mobile handoff design since a mobile at 20 GHz would see approximately 21 dB of fading when moving around a marble corner from a diffraction angle of 0$^\circ$ to 20$^\circ$. For a person moving at a speed of 1 m/s, this results in about a 21 dB/s initial fade rate from LOS to NLOS. Fig. \ref{fig:Outdoor} shows more oscillation patterns in the marble corner measurements in the deeply shadowed region than in the stone pillar measurements, which indicates more prevalent scattering when measuring the marble corner. \textcolor{black}{We note that the diffraction loss for outdoor building corners at mmWaves using directional antennas can be better predicted with a simple linear model (creeping wave), whereas the KED model agrees well with indoor diffraction loss measurements. Both the creeping wave and KED models can be used in network simulations and ray-tracers with short computation time and good accuracy while considering approximately 5-6 dB standard deviation (see~\cite{Deng16a} for mean error and standard deviation values between the measured data and models derived from the data).} For cross-polarized diffraction measurement results, see~\cite{Deng16b}.

\section{Human Blockage Measurements and Models}
\subsection{Introduction of mmWave Human Blockage}
In mmWave communications, attenuation caused by human blockage (when a human body blocks the LOS path between a transmitter and receiver) will greatly impact cellphone link performance, and phased array antennas will need to adapt to find other propagation paths when blocked by a human~\cite{Friis46a, Mac16a}. This is in sharp contrast to omnidirectional antennas used at sub-6 GHz frequencies. Understanding this severe blockage effect and employing appropriate models for mobile system simulation are important for properly designing future mmWave antennas and beam steering algorithms~\cite{Sun15a,Samimi16a,Sun14b}. One of the earliest human blockage measurement studies~\cite{Sato98a} was conducted at 60 GHz for indoor wireless local area networks (WLAN) with T-R separation distances of 10 m or less in a typical office environment. Results showed that the signal level decreased by as much as 20 dB when a person blocked the direct path between omnidirectional TX and RX antennas, with deep fades reaching 30 dB using directive antennas\cite{Collonge04a},~\cite{Collonge03a}. In addition to human blockage measurements at 60 GHz,~\cite{80211ad10a} provided a human induced cluster blockage model based on ray tracing~\cite{Seidel94a}, a random walk model, and a diffraction model from contributions to 802.11ad~\cite{Jacob09d},~\cite{Jacob09e}. The probability distributions for four parameters (duration, decay time, rise time, and mean attenuation) were generated by the human-induced cluster blockage model and were validated with the Kolmogorov-Smirnov test~\cite{80211ad10a}. The \textit{mobile and wireless communications enablers for the twenty-twenty information society (METIS)} and \textit{3\textsuperscript{rd} Generation Partnership Project (3GPP)} also proposed their own human blockage models~\cite{METIS15a,3GPP.38.900,Medbo13a} based on KED models for one or multiple edges. In the following subsections, 73 GHz human blockage measurements and an improved human blockage model are described.

\subsection{Human Blockage Measurement System}
A real-time spread spectrum correlator channel sounder system described in~\cite{Mac16a} was used for the human blockage measurements. A pseudorandom noise (PN) sequence of length 2047 was generated at baseband with a field programmable gate array (FPGA) and high-speed digital-to-analog converter (DAC). This wideband sequence was modulated to an intermediate-frequency (IF) of 5.625 GHz which was then upconverted to a center frequency of 73.5 GHz (1 GHz null-to-null RF bandwidth). The transmit power at the TX was -5.8 dBm, and identical TX and RX horn antennas with 15$^\circ$ azimuth and elevation (Az./El.) HPBW and 20 dBi of gain were used. The received signal was downconverted to IF, and then demodulated to its in-phase ($I$) and quadrature ($Q$) baseband signals that were sampled at 1.5 Giga-Samples (GS/s) via a high-speed analog-to-digital converter (ADC). The digital signals were then correlated in software via a Fast Fourier Transform (FFT) matched filter to create the $I$ and $Q$ channel impulse response (CIR), and subsequent power delay profiles (PDPs) ($I^2+Q^2$). The system had a multipath resolution of 2 ns and an instantaneous dynamic range of 40 dB and could capture PDPs with a minimum consecutive snapshot interval of 32.752$\mu s$ to measure rapid fading. Power was computed as the area under the PDP, and the voltage was found as the square root of power~\cite{Rap02a}. 

\subsection{Blockage Measurement Environment Description}
\begin{figure}[tb!]
	\centering
	\includegraphics [width=0.45\textwidth]{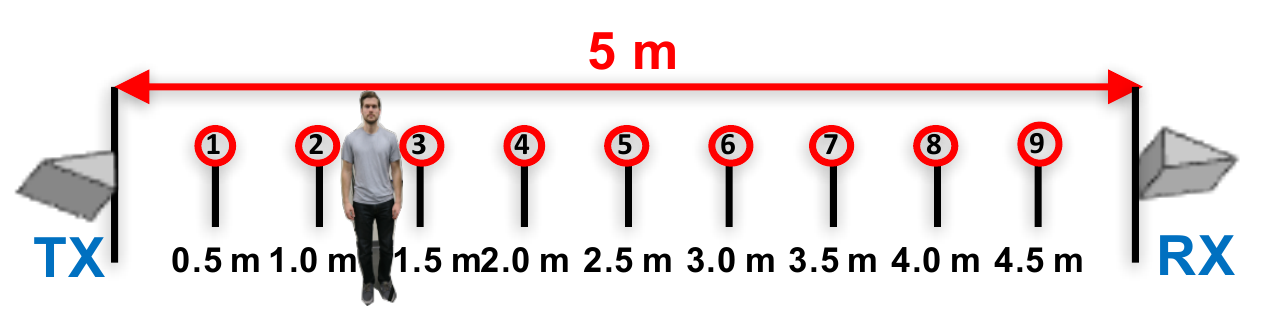}
	\caption{Depiction of nine measurement locations where at each indicator separated by 0.5 m, the human blocker walked at a perpendicular orientation between the TX and RX antennas~\cite{Mac16a}.}\label{fig:4}
	\squeezeup
\end{figure}

Human blockage measurements were conducted in an open laboratory using a 5 m T-R separation distance. High gain narrowbeam horn antennas were used at the TX and RX with both antenna heights set to 1.4 m relative to the ground and aligned on boresight. Nine measurements were recorded with a human blocker walking at a perpendicular orientation through the LOS path between the TX and RX at an approximate 1 m/s speed. This perpendicular walk was performed at 0.5 m increments between the TX and RX starting at 0.5 m from the TX for measurement one, and 4.5 m from the TX for measurement nine, as depicted in Fig. \ref{fig:4}. We note that 0.5 m is the typical distance for a person to view the screen of a smartphone. For each of the nine perpendicular walks, 500 PDPs were recorded per second in a five-second window, resulting in 2500 PDPs for each measurement. The dimensions of the human blocker were: $b_{\text{breadth}}=0.47$ m; $b_{\text{depth}}=0.28$ m; $b_{\text{height}}=1.80$ m. Detailed information about the experiment is given in~\cite{Mac16a,Mac17a}.

\subsection{KED Blockage Model}
\begin{figure}
	\centering
	\begin{subfigure}[b]{0.38\textwidth}
		\includegraphics[width=1\linewidth]{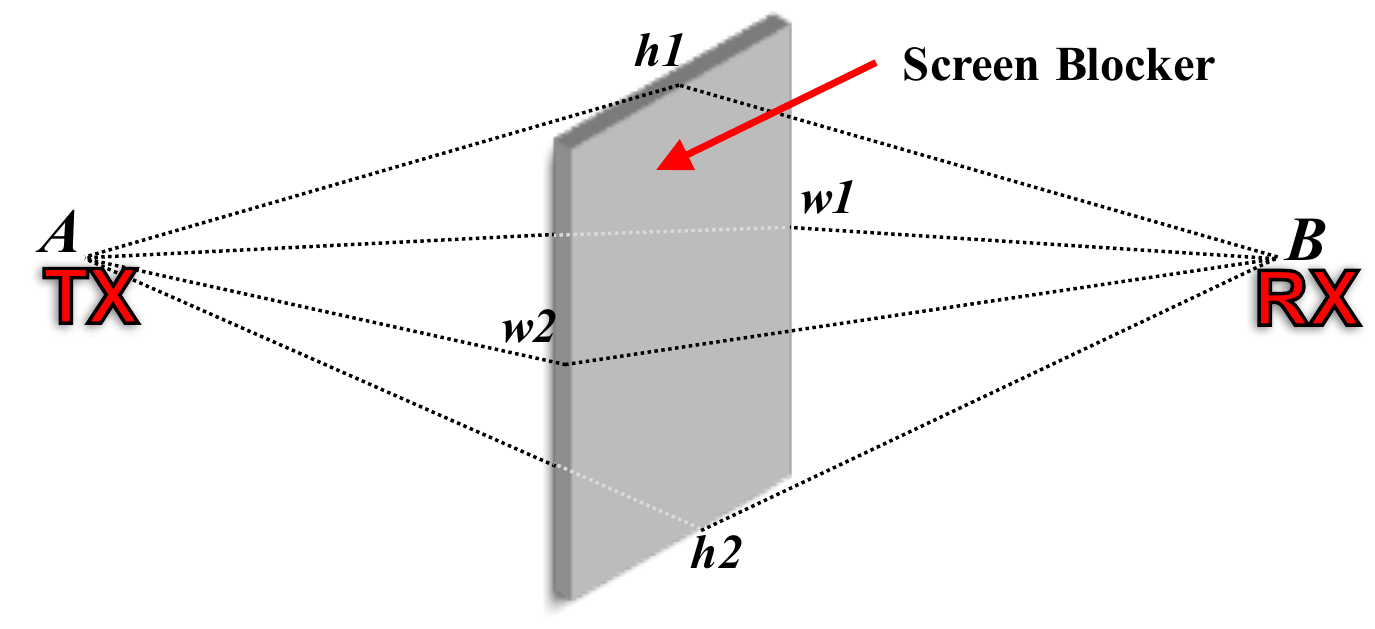}
		\caption{3D screen projection.}
		\label{fig:DKEDmeasDProj} 
	\end{subfigure}
	\begin{subfigure}[b]{0.38\textwidth}
		\includegraphics[width=1\linewidth]{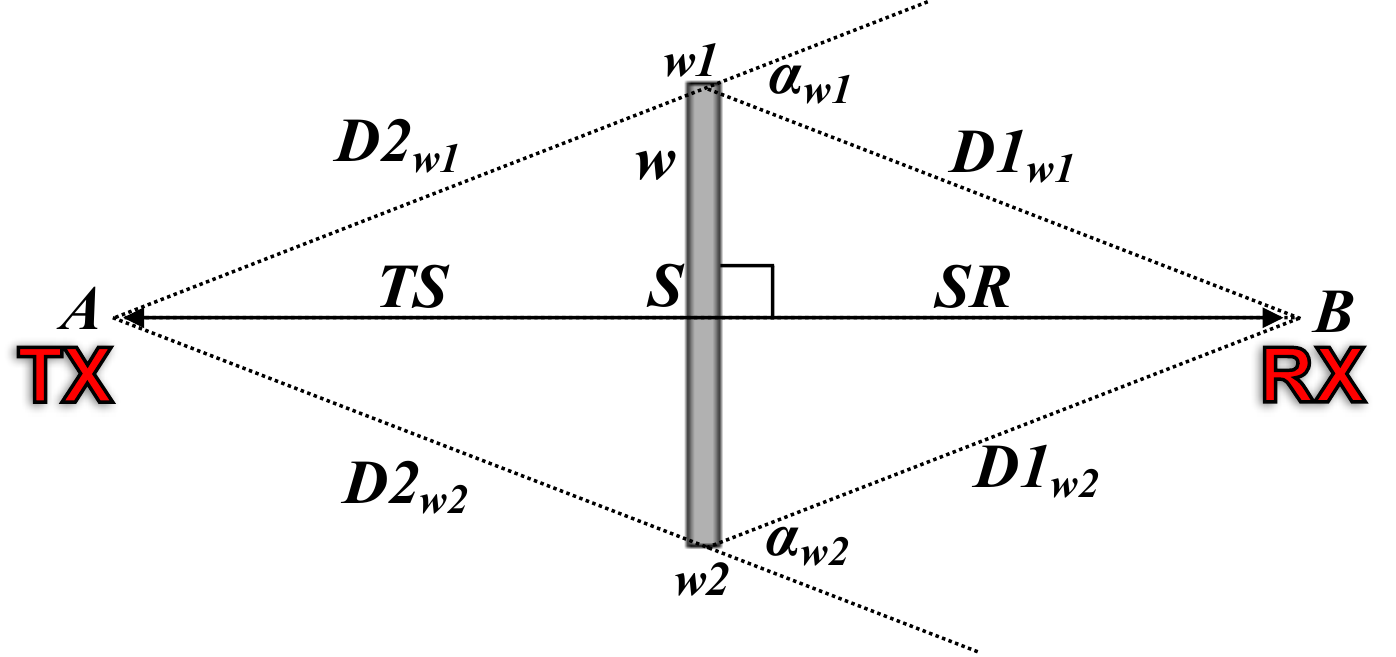}
		\caption{Top view of screen projection.}
		\label{fig:TopProj}
	\end{subfigure}
	\caption{(a) 3D and (b) top-down projection of screen blocker.}\label{fig:2} 
	\squeezeup
\end{figure}

Knife-edge diffraction is commonly used to model human blockage by modeling a thin rectangular screen as the blocker~\cite{Kunisch08a, Medbo13a,METIS15a,3GPP.38.900}. In DKED modeling, the rectangular screen is considered infinitely high, such that diffraction loss only occurs from the two side edges of the body. A typical screen blocker with height $h$ and width $w$ is displayed in Fig. \ref{fig:2} from both a 3D and top view screen projection. The dimensions for the top view of the screen are defined as follows:  $w$ is the width of the screen from $w1$ to $w2$; $r\overset{\text{def}}{=}\overline{AB}$; $w\overset{\text{def}}{=}\overline{w1w2}$; $h\overset{\text{def}}{=}\overline{h1h2}$; $TS\overset{\text{def}}{=}\overline{AS}$;
$SR\overset{\text{def}}{=}\overline{SB}$; $D2_{w1}\overset{\text{def}}{=}\overline{A w1}$; $D1_{w1}\overset{\text{def}}{=}\overline{w1 B}$; $D2_{w2}\overset{\text{def}}{=}\overline{A w2}$; $D1_{w2}\overset{\text{def}}{=}\overline{w2 B}$; $\alpha_{w1}$ and $\alpha_{w2}$ are the diffraction angles for the $w1$ and $w2$ edges of the screen, respectively~\cite{Mac16a}.
From Fig. \ref{fig:TopProj}, the two side edges of the screen are denoted $w1$ and $w2$, where the distance between the edges is the body depth ($b_{\text{depth}}$) or width of the screen $w$, since the blocker walks through the LOS path at a perpendicular orientation. As the screen moves between the TX and RX antennas and blocks the LOS path, the screen is always considered perpendicular to the solid line drawn between the two, to reduce computational complexity.

Diffraction loss is calculated via numerical approximation by Fresnel integration and the diffraction parameter as follows~\cite{Wang15a}:
\begin{equation}\label{eq:KEDapprox}
\footnotesize
F_{w1|w2}=
\begin{cases}
\frac{(1-j)}{2}\left(\frac{1+j}{2}-(C(v)+j\cdot S(v))\right)\text{, if $v>$ 0} \\ 
\frac{(1-j)}{2}\left(\frac{1+j}{2}+(C(-v)+j\cdot S(-v))\right)\text{, otherwise}
\end{cases}
\end{equation}
where the numerical approximations of Fresnel integration for $C(v)$ and $S(v)$ are:
\begin{subequations}
	\footnotesize
	\begin{equation}\label{eq:FresC}
	C(v) = \int_{0}^{v}\cos{\left(\frac{\pi v^2}{2}\right)dv}
	\end{equation}
	\begin{equation}\label{eq:FresC}
	S(v) = \int_{0}^{v}\sin{\left(\frac{\pi v^2}{2}\right)dv}
	\end{equation}
\end{subequations}
and where the diffraction parameter $v$ is derived by~\cite{Rap02a}:
\begin{equation}\label{eq:diffp}
\footnotesize
v_{w1|w2} = \pm\alpha_{w1|w2} \sqrt{\frac{2\cdot \overline{AS}\cdot \overline{SB}}{\lambda (\overline{AS}+\overline{SB})}}
\end{equation}
The diffraction parameter $v$ is calculated based on the distance from the TX to the screen, from the screen to the RX, the diffraction angle $\alpha$ (see Fig.~\ref{fig:TopProj}), and the carrier wavelength $\lambda$. The $\pm$ sign in~\eqref{eq:diffp} is applied as $+$ to both edges for NLOS conditions. \textcolor{black}{When calculating the diffraction parameter~\eqref{eq:diffp} under unobstructed (LOS) conditions for the screen edge ($w1$ or $w2$) closest to the straight line drawn between the TX and RX, the $\pm$ is treated as ``$-$", whereas the $\pm$ is treated as ``$+$" for the screen edge farthest from the straight line drawn between the TX and RX~\cite{Wang15a}}. 

The individual received signal caused by knife-edge diffraction from the $w1$ and $w2$ edges is $F_{w1}$ and $F_{w2}$, respectively. The complex signals corresponding to the edges can be added in order to determine the combined diffraction loss observed at the RX. The total diffraction loss power in log-scale is determined by taking the magnitude squared of the summed signals as follows~\cite{Wang15a}: 
\begin{equation}\label{eq:Lscreen}
\footnotesize
L_{\text{screen}}[\dB] = 20\log_{10} \left( \big \lvert F_{w1}+F_{w2} \big \rvert \right)
\end{equation}
The DKED model~\eqref{eq:Lscreen} has been adopted by METIS and others~\cite{METIS15a,3GPP.38.900}, but it does not consider the antenna radiation pattern (it assumes an omnidirectional antenna~\cite{Medbo04a}) and has been shown to underestimate diffraction loss in the deepest fades when using directional antennas, which are sure to be employed by mmWave mobile devices~\cite{Mac16a}. This physical phenomenon occurs when the blocker obstructs the LOS path between antennas (i.e. deep fading), leaving only the off-boresight antenna gains to contribute to the received signal strength, which is slightly less than the directive gain. To account for the impact of non-uniform gain directional antennas on human blockage, antenna gain is considered in~\eqref{eq:Lscreen} (see~\cite{Mac16a}). The following azimuth far-field power radiation pattern of a horn antenna (general for any directional antenna) for a given half-power beamwdith (HPBW) is approximated by~\cite{Sun15a,Mac16a}:
\vspace{0.2cm}
\begin{align*}
\footnotesize
G(\theta) = \sinc^2(\textrm{a}\cdot \sin (\theta))\cdot \cos^2(\theta)
\end{align*}
where:
\begin{align*}
\footnotesize
\sinc^2\Bigg(\textrm{a}\cdot \sin \bigg( \frac{\text{HPBW}_{\text{AZ}}}{2}\bigg)\Bigg)\cdot \cos^2 \bigg( \frac{\text{HPBW}_{\text{AZ}}}{2}\bigg)=\frac{1}{2}
\end{align*}
The DKED model in~\eqref{eq:Lscreen} can be extended to include TX and RX antenna gains for the projected angles $\theta$ between the TX and the screen, and the screen and RX as follows~\cite{Mac16a}:
\begin{equation}\label{eq:LscreenMod}
\footnotesize
\begin{split}
L_{\textrm{Screen A.G.}}[\dB] = 20\log_{10} \Bigg( \Bigg \lvert F_{w1}\cdot \sqrt{G_{D2_{w1}}}\cdot \sqrt{G_{D1_{w1}}} \\+F_{w2} \cdot \sqrt{G_{D2_{w2}}}\cdot \sqrt{G_{D1_{w2}}}\Bigg \rvert \Bigg)
\end{split}
\end{equation}
where $G_{D2_{w1}},\;G_{D1_{w1}},\;G_{D2_{w2}}$, and $G_{D1_{w2}}$ are the linear power gains (normalized to the directive gain such that $G_{0^\circ}=1$) of the antennas based on the point-source projections $\overline{Aw1}$; $\overline{w1B}$; $\overline{Aw2}$; $\overline{w2B}$; $A$ to $w1$, $w1$ to $B$, $A$ to $w2$, and $w2$ to $B$ (see Fig.~\ref{fig:TopProj}). When the screen does not obstruct the LOS path between the TX and RX, the normalized gains are set to $G(\theta)=1$, since the slight variations of antenna patterns have little effect on diffraction loss in the unobstructed case.

\subsection{Human Blockage Results and Analysis}
\begin{figure}[tb!]
	\squeezeup
	\begin{center}
		\includegraphics [width=0.5\textwidth]{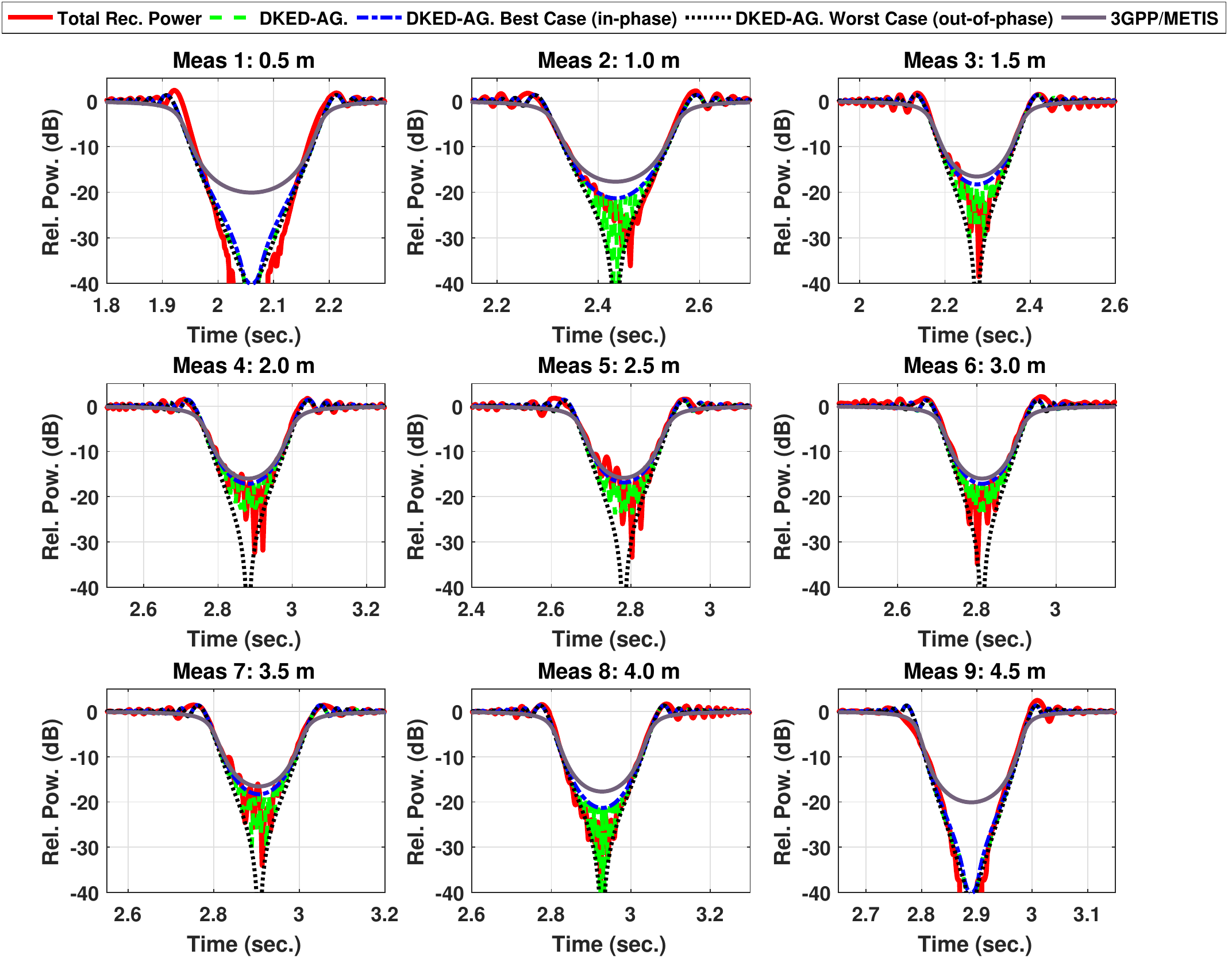}
		\caption{Comparison of measured received power of human blockage at 73 GHz and the DKED-AG model in~\eqref{eq:LscreenMod}~\cite{Mac16a}.}\label{fig:DKEDmeas}
	\end{center}
	\squeezeup
\end{figure}

For each of the nine measurement paths, the area under the curve of each of the 2500 PDPs was integrated to calculate the received power in 2 ms increments. In Fig. \ref{fig:DKEDmeas} the received power (red) is compared to the DKED antenna gain (DKED-AG) model (green)~\eqref{eq:LscreenMod}, in addition to showing the constructive (signals in-phase) and destructive (signals out of phase) sum of received signals of the upper (blue) and lower (black) bound of the fade envelope, respectively. 
Fig. \ref{fig:DKEDmeas} represents loss as compared to a free space reference with no blockage between the TX and RX. From Fig. \ref{fig:DKEDmeas}, we observe gain in the received signal as the human enters the TX/RX LOS path, and then deep attenuations as the human blocks the LOS path. Due to the fact that identical antennas were used at the TX and RX, the envelopes of the received signal power were similar in the two different symmetrical cases (i.e., Meas. 1 and Meas. 9). The best case scenario  (minimum diffraction loss) is found by summing the magnitudes of received field components from the $w1$ and $w2$ edges of the blocker, represented by the blue dashed line in Fig.~\ref{fig:DKEDmeas}. For minimum loss,~\eqref{eq:LscreenMod} is reformulated as: $20\log_{10}(|F_{w1}\cdot \sqrt{G_{D2_{w1}}}\cdot \sqrt{G_{D1_{w1}}}|+|F_{w2} \cdot \sqrt{G_{D2_{w2}}}\cdot \sqrt{G_{D1_{w2}}}|)$. The worst case scenario (maximum diffraction loss) is found by taking the difference of the magnitudes of received signals from the $w1$ and $w2$ edges, and is represented by the black dotted line in Fig~\ref{fig:DKEDmeas}, where~\eqref{eq:LscreenMod} is computed as: $20\log_{10}\Big(\big||F_{w1}\cdot \sqrt{G_{D2_{w1}}}\cdot \sqrt{G_{D1_{w1}}}|-|F_{w2} \cdot \sqrt{G_{D2_{w2}}}\cdot \sqrt{G_{D1_{w2}}}|\big|\Big)$.

It was previously demonstrated that diffraction loss models that do not account for antenna gain pattern can severely underestimate the diffraction loss when the blocker is close to either antenna (a critical issue for mobile phone use)~\cite{Mac16a}. The DKED-AG model~\eqref{eq:LscreenMod} accurately predicts what is measured, and predicts the deepest attenuation caused by a human blocker in excess of 40 dB. To model multiple blockers, the screen model can be replicated multiple times. These results show that adaptive antenna array and beamforming techniques will be employed to find suitable reflectors and \textcolor{black}{scatterers} in the signal transmission to overcome severe blockage attenuation in future 5G communication systems. The DKED-AG model in~\eqref{eq:LscreenMod} may be extended~\cite{Kunisch08a} to consider the top and bottom screen edges, phase corrections, and non-perpendicular screen orientations, although the simple model~\eqref{eq:LscreenMod} matches the human blocking measurements with confidence. It can be seen in Fig. \ref{fig:DKEDmeas} (Meas. 1 and 9) that the signal strength drops off at a rate of 0.4 dB/ms as the blocker \textcolor{black}{moves at 1 m/s and begins to} shadow the TX (RX). Mobile handoffs and beam steering schemes will be needed to rescue the mobile from severe fades by the use of electrically scanning beams at the sub-millisecond level, a feat easily accomplished with sub millisecond packets in an air interface standard. \textcolor{black}{An additional technique for mitigating the effects of rapid fading could include rapid re-routing around obstacles via handoff to another access point (AP) in a network cluster~\cite{Ghosh16a}.} \textcolor{black}{Note that just prior to the deep shadowing events in Fig.~\ref{fig:DKEDmeas} there is a slight increase/scintillation of signal strength of $\sim$ 2 dB peak-to-peak amplitudes (noticed by others in~\cite{Jacob10b}), which could be used to detect the imminent presence of an obstruction such that the RX adapts its beam in anticipation of the pending deep fade. The 3GPP/METIS blockage model~\cite{3GPP.38.900,METIS15a} shown in Fig.~\ref{fig:DKEDmeas} underestimates the deep fades of shadowing events~\cite{Mac16a}, especially when the blocker is close to the TX or RX antenna, since the full directive gain of the TX and RX antennas is not available across the diffraction obstacle, and thus is unable to contribute to the received signal strength from diffraction around the blocker during the shadowing event~\cite{Mac16a}. We note that the 3GPP/METIS model only offers reasonable agreement to the measured loss when the blocker is far (several meters) from the TX and RX antenna.}

\section{Small-Scale Spatial Statistics}\label{sec:smallscale}
\subsection{Introduction of Small-Scale Spatial Statistics}
Small-scale fading and small-scale autocorrelation characteristics are crucial for the design of future mmWave communication systems, especially in multiple-input multiple-output (MIMO) channel modeling. Previous studies on small-scale fading characteristics focused on sub-6 GHz frequencies, yet investigations at mmWave are scarce. Wang \textit{et al.}~\cite{Wang15b} showed that small-scale fading of received power in indoor corridor scenarios with omnidirectional antennas at both TX and RX could be well described by Ricean distributions with K-factors ranging from 5 dB to 10 dB based on their indoor corridor measurements at 15 GHz with a bandwidth of 1 GHz, and ray tracing results using a ray-optical based channel model validated by measurements. Henderson \textit{et al.} compared Rayleigh, Ricean, and the Two-Wave-Diffuse-Power (TWDP) distributions to find the proper small-scale fading distribution of received voltage magnitudes for a measured 2.4 GHz indoor channel~\cite{Henderson08a} where the Ricean distribution had highest modeling accuracy in most indoor cases~\cite{Henderson08a}. The authors in~\cite{Romero-Jerez16a} demonstrated the use of the TWDP fading model for mmWave communications. It was reported that \textcolor{black}{log-normal} distribution had a good fit to measured received signal envelopes in some indoor mobile radio channels~\cite{Cotton07a}. Important work on wideband directional small-scale fading also appears in~\cite{Holtzman94a,Durgin03a,Sun17a,Samimi16c,Dupleich17a}. In the following subsections, small-scale fading distributions of total power and autocorrelation characteristics of received voltage amplitudes at 73 GHz in urban microcell environments are investigated based on a measurement campaign conducted during the summer of 2016 around the engineering campus of New York University in downtown Brooklyn. 

\subsection{Measurement System for Small-Scale Spatial Statistics}
The TX system for small-scale fading and autocorrelation measurements at 73 GHz was identical to the TX system used for the human blocking measurements with the difference only for TX antennas and transmit powers as identified in Table ~\ref{tbl:2}. The RX side of the system captured the RF signal via steerable horn antennas and downconverted the signal to an IF of 5.625 GHz, which was then demodulated into its baseband in-phase ($I$) and quadrature-phase ($Q$) signals which were correlated via a common sliding correlation architecture~\cite{Rap02a,Rap15b,Mac15b,Rap13a} where the time-dilated $I$ and $Q$ channel voltages were sampled by an oscilloscope and then squared and added together in software to generate a PDP. Antennas with 27 dBi gain (7$^\circ$ Az./El. HPBW) and 9.1 dBi gain (60$^\circ$ Az./El.HPBW) were used at the TX and RX sides, respectively\textcolor{black}{\cite{Sun17a}}.

\begin{table}[tb!]
	\renewcommand{\arraystretch}{1.4}
	\caption{Hardware Specifications of Small-Scale Fading and Local Area Channel Transition Measurements.}~\label{tbl:2}
	\fontsize{7.0}{7.0}\selectfont
	\begin{center}
		\squeezeup
		\begin{tabular}{|>{\centering\arraybackslash}m{3.0cm}|>{\centering\arraybackslash}m{2.2cm}|>{\centering\arraybackslash}m{2.2cm}|>{\centering\arraybackslash}m{2.5cm}|>{\centering\arraybackslash}m{2.5cm}|>{\centering\arraybackslash}m{0.6cm}|}\hline
			\textbf{Campaign}	                        & \textbf{73 GHz Small-Scale Fading and Correlation Measurements}	        & \textbf{73 GHz Local Area Channel Transition Measurements}	               \\ \hline \hline
			\textbf{Broadcast Sequence}			        & \multicolumn{2}{c|}{$11\textsuperscript{th}$ order PN Code (L = $2^{11}-1$ = 2047)}   \\ \hline
			\textbf{TX and RX Antenna Type}             & \multicolumn{2}{c|}{Rotatable pyramidal horn antenna}  \\ \hline
			\textbf{TX/RX Chip Rate} 			            & \multicolumn{2}{c|}{500 Mcps / 499.9375 Mcps}		            \\ \hline
			\textbf{Slide Factor $\gamma$}              & \multicolumn{2}{c|}{8000}                     \\ \hline
			\textcolor{black}{\textbf{RF Null-to-Null Bandwidth}} 	    & \multicolumn{2}{c|}{\textcolor{black}{1 GHz}}			           \\ \hline
			\textbf{PDP Threshold}                      & \multicolumn{2}{c|}{20 dB down from max peak}  \\ \hline
			\textbf{TX/RX Intermediate Freq.}       & \multicolumn{2}{c|}{5.625 GHz}          \\ \hline
			\textbf{TX/RX Local Oscillator}             & \multicolumn{2}{c|}{67.875 GHz (22.625 GHz $\times$ 3)}   \\ \hline
			\textbf{Carrier Frequency}                  & \multicolumn{2}{c|}{73.5 GHz}        \\ \hline
			\textbf{TX Antenna Gain}                    & \multicolumn{2}{c|}{27 dBi}  \\ \hline
			\textbf{RX Antenna Gain}                    & 9.1 dBi   & 20 dBi \\ \hline
			\textbf{Max TX Power / EIRP}                           & 14.2 dBm / 41.2 dBm   & 14.3 dBm / 41.3 dBm   \\ \hline
			\textbf{TX Az. and El. HPBW}      & \multicolumn{2}{c|}{$7^\circ$}  \\ \hline
			\textbf{TX/RX Heights}                         & 4.0 m / 1.4 m & 4.0 m / 1.5 m   \\ \hline
			\textbf{RX Az. and El. HPBW}      & $60^\circ$  & $15^\circ$ \\ \hline
			\textbf{TX-RX Antenna Pol.}         & \multicolumn{2}{c|}{V-V (vertical-to-vertical)}   \\ \hline
			\textbf{Max Measurable Path Loss}       & 168 dB   & 180 dB     \\ \hline
		\end{tabular}
	\end{center}
\end{table}

\subsection{Small-Scale Measurement Environment and Procedure}
\begin{figure}
	\centering
	\includegraphics[width=2.9in]{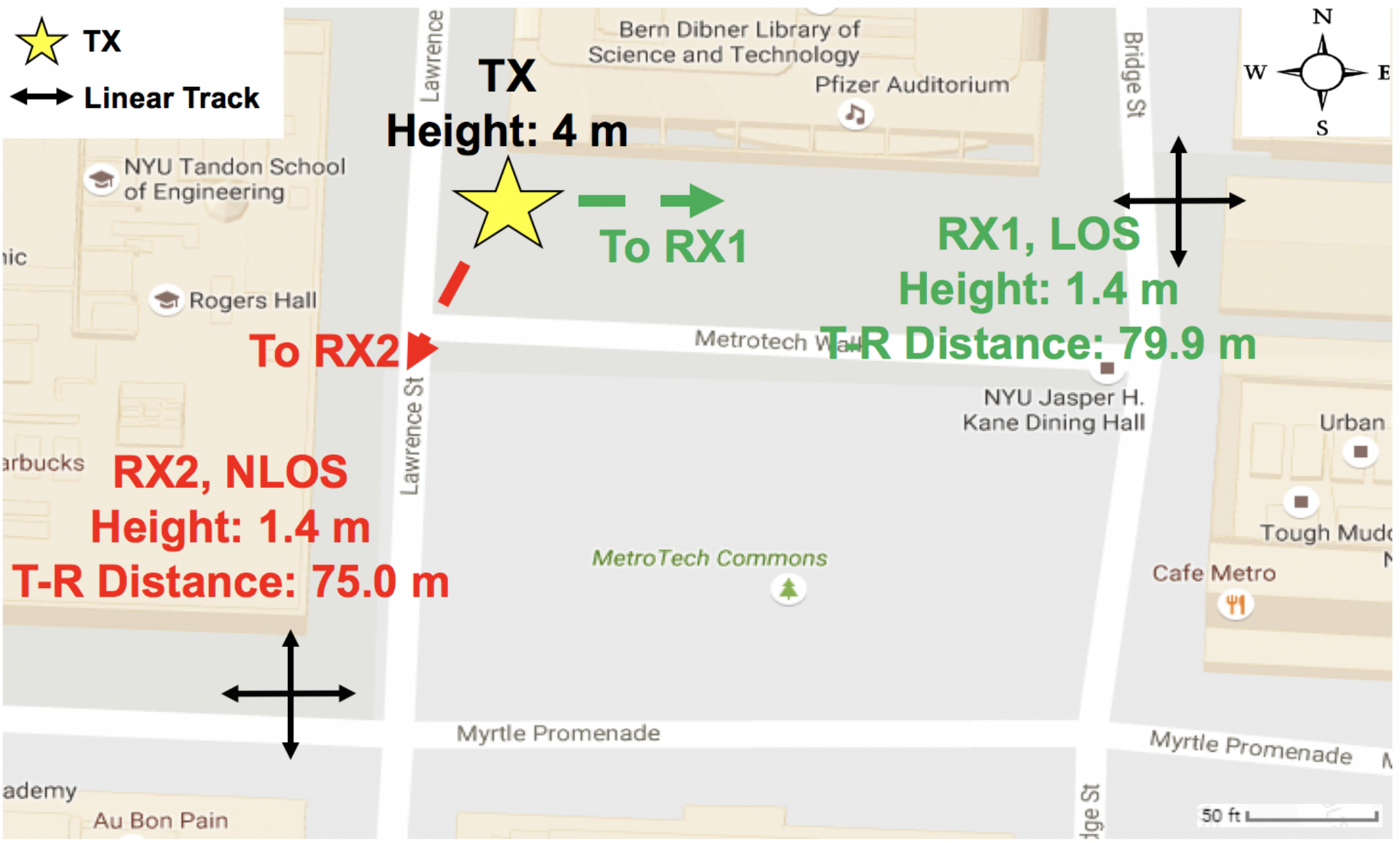}
	\caption{\textcolor{black}{2D map of the 73 GHz small-scale measurement environment and the locations of TX and RX. Pointing to the top of the map is 0$^\circ$.}}
	\label{fig:TX_RX_Location}
	\squeezeup
\end{figure}

In the summer of 2016, a set of small-scale linear track measurements were conducted at 73 GHz on the campus of NYU Tandon School of Engineering in downtown Brooklyn, New York, representative of an urban microcell (UMi) environment\textcolor{black}{\cite{Sun17a}}. The measurement environment, and the TX and RX locations are depicted in Fig.~\ref{fig:TX_RX_Location}. One TX location with an antenna height of 4.0 m above the ground and two RX locations with an antenna height of 1.4 m were selected to perform the measurements, where one RX was LOS to the TX while the other was NLOS. The TX was placed near the southwest corner of the Dibner library building (top center in Fig.~\ref{fig:TX_RX_Location}), the LOS RX was located 79.9 m away from the TX, and the NLOS RX was shadowed by the southeast corner of a building (Rogers Hall on the map) with a T-R separation distance of 75.0 m\textcolor{black}{\cite{Sun17a}}. Other specifications about the measurement hardware are detailed in Table~\ref{tbl:2}. 

A fixed 35.31-cm spatial linear track \textcolor{black}{(about 87 wavelengths at 73.5 GHz)} was used at each RX location in the small-scale fading measurements, over which the RX antenna was moved in increments of half-wavelength (2.04 mm) over 175 track positions\textcolor{black}{\cite{Sun17a}}. At each RX, six sets of small-scale fading measurements were performed, where the elevation angle of the RX antenna remained fixed at 0$^\circ$ (parallel to the horizon) and a different azimuth angle was chosen for each set of the measurements with the adjacent azimuth angles separated by 60$^\circ$, such that the RX antenna swept over the entire azimuth plane after rotating through the six pointing angles. The RX antenna was pointing at a fixed angle while moving along the linear track for each set of the measurements and a PDP was acquired at each track position for each pointing angle. The TX antenna elevation angle was always fixed at 0$^\circ$ (parallel to horizon). Under the LOS condition, the TX antenna was pointed at 90$^\circ$ in the azimuth plane, directly towards the RX location; for NLOS, the TX antenna azimuth pointing angle was 200$^\circ$, roughly towards the southeast corner of Rogers Hall in Fig.~\ref{fig:TX_RX_Location}\textcolor{black}{\cite{Sun17a}. Due to space limitations, we show here only one track orientation at each RX (along the direction of the street beside the RX), but more results and observations are detailed in~\cite{Sun17a} which show the fading depths are a function of antenna orientations and environment.}

As a comparison, the 28 GHz small-scale measurements presented in~\cite{Samimi16b} investigated the small-scale fading and autocorrelation of \textit{individual resolvable multipath} voltage amplitudes using a 30$^\circ$ Az./El. HPBW RX antenna, whereas this paper studies 73 GHz fading and autocorrelation using a wider HPBW (60$^\circ$) RX antenna, and focuses on received signal voltage amplitude by integrating the area under the \textit{entire PDP} curve and then taking the square root of the total power, instead of individual multipath voltage amplitude at each location along a track.

\begin{figure}
	\centering
	\includegraphics[width=3.2in]{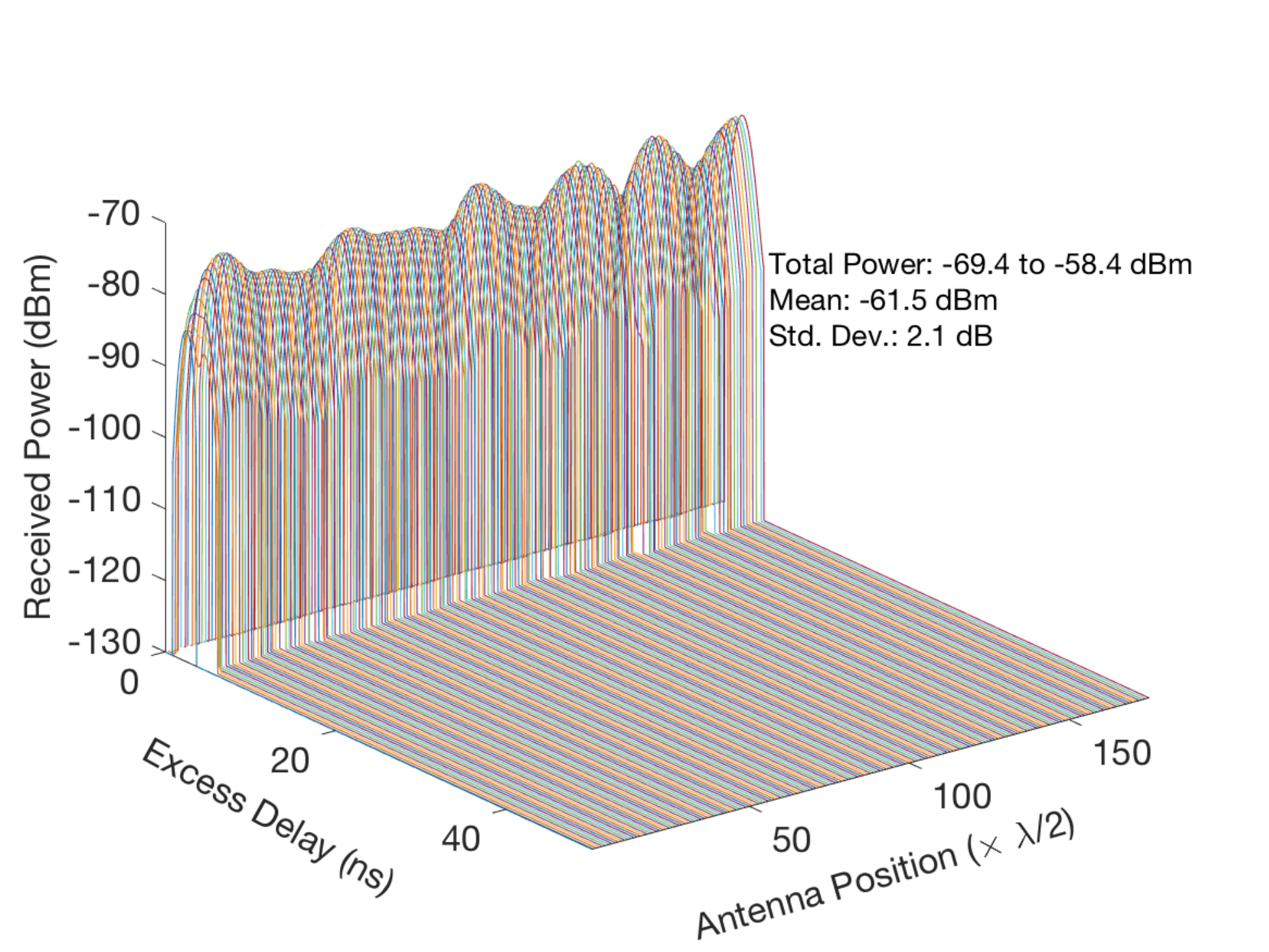}
	\caption{\textcolor{black}{Measured 73 GHz small-scale directional PDPs over 175 track positions in LOS. The RX horn antenna (60$^\circ$ HPBW) was pointing on boresight to the TX, and track orientation orthogonal to the T-R line.}}
	\label{fig:LOSPDP2}
	\squeezeup
\end{figure}

\begin{figure}
	\centering
	\includegraphics[width=3.2in]{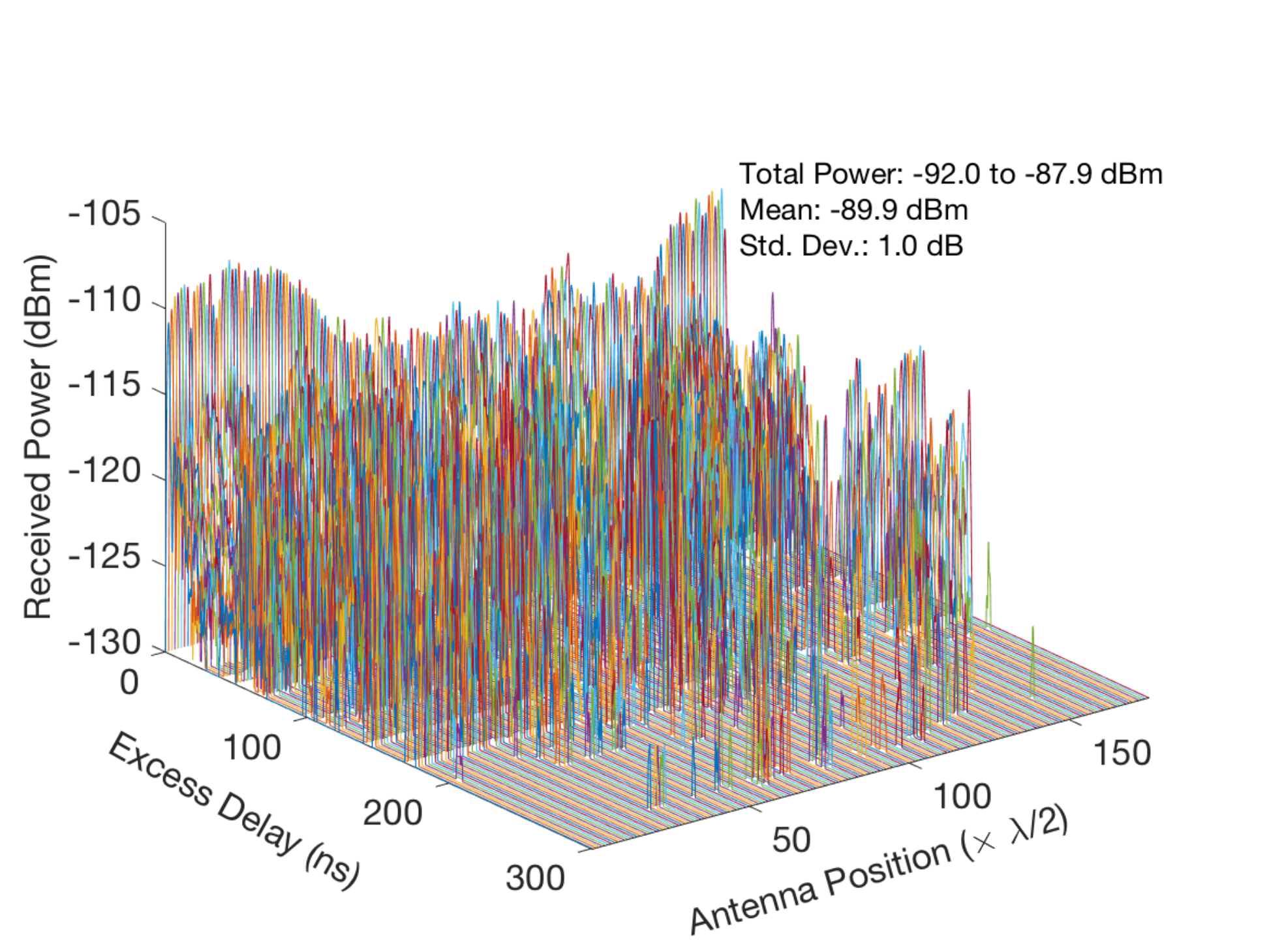}
	\caption{\textcolor{black}{Measured 73 GHz small-scale directional PDPs over 175 track positions in NLOS. The RX horn antenna (60$^\circ$ HPBW) was pointing to the TX but was obstructed by a building corner, and the track orientation was along the direction of the street.}}
	\label{fig:NLOSPDP2}
\end{figure}

\textcolor{black}{\subsection{Small-Scale Measurement Results}}
\textcolor{black}{Fig.~\ref{fig:LOSPDP2} illustrates typical measured small-scale directional PDPs over 175 track positions on the 35.31-cm (about 87 wavelengths at 73.5 GHz) linear track in the LOS environment, where the RX horn antenna with 60$^\circ$ HPBW was pointing on boresight to the TX, and the track orientation was orthogonal to the T-R line. The total power in Figs.~\ref{fig:LOSPDP2} and~\ref{fig:NLOSPDP2} is computed as the area under the PDP at a particular track position over the 1 GHz RF bandwidth. Fig.~\ref{fig:LOSPDP2} shows there is 11 dB power variation over different track positions, but the power variation is only 3.7 dB when the track orientation was in the direction of the T-R line (not shown), indicating little small-scale spatial fading~\cite{Sun17a}.}

\textcolor{black}{Typical measured small-scale directional PDPs over 175 track positions on the 35.31-cm (about 87 wavelengths at 73.5 GHz) linear track in the NLOS environment are depicted in Fig.~\ref{fig:NLOSPDP2}, where the track orientation was along the direction of the street, and the RX antenna was pointing to the TX but was obstructed by a building corner~\cite{Sun17a}. Fig.~\ref{fig:NLOSPDP2} shows there is very moderate power variation (4.1 dB) over different local-area track positions, albeit with rich and varying multipath components.}\\


\subsection{Small-Scale Spatial Statistics Results and Analysis}
\subsubsection{Omnidirectional Small-Scale Spatial Statistics}
As described above, a rotatable directive horn antenna was used at the RX side to capture directional PDPs in the small-scale fading and correlation measurements. In channel modeling, however, omnidirectional statistics are often preferred, since arbitrary antenna patterns can be implemented according to one's own needs if accurate temporal and spatial statistics are known~\cite{Sun17c}. Therefore, we synthesized the approximated omnidirectional received power at every track interval by taking the \textcolor{black}{area under the curve of each directional PDP and summing powers using the approach presented in~\cite{Sun15a} and on Page 3040 from~\cite{Rap15b}}, thereby computing omnidirectional received power. Although the RX antenna did not sweep the entire 4$\pi$ Steradian sphere, the azimuth plane spanned $\pm 30^{\circ}$ with respect to the horizon, ensuring that a large majority of the arriving energy was captured, as verified in~\cite{Sun15a}. 

\begin{figure}
	\squeezeup
	\centering
	\includegraphics[width=2.7in]{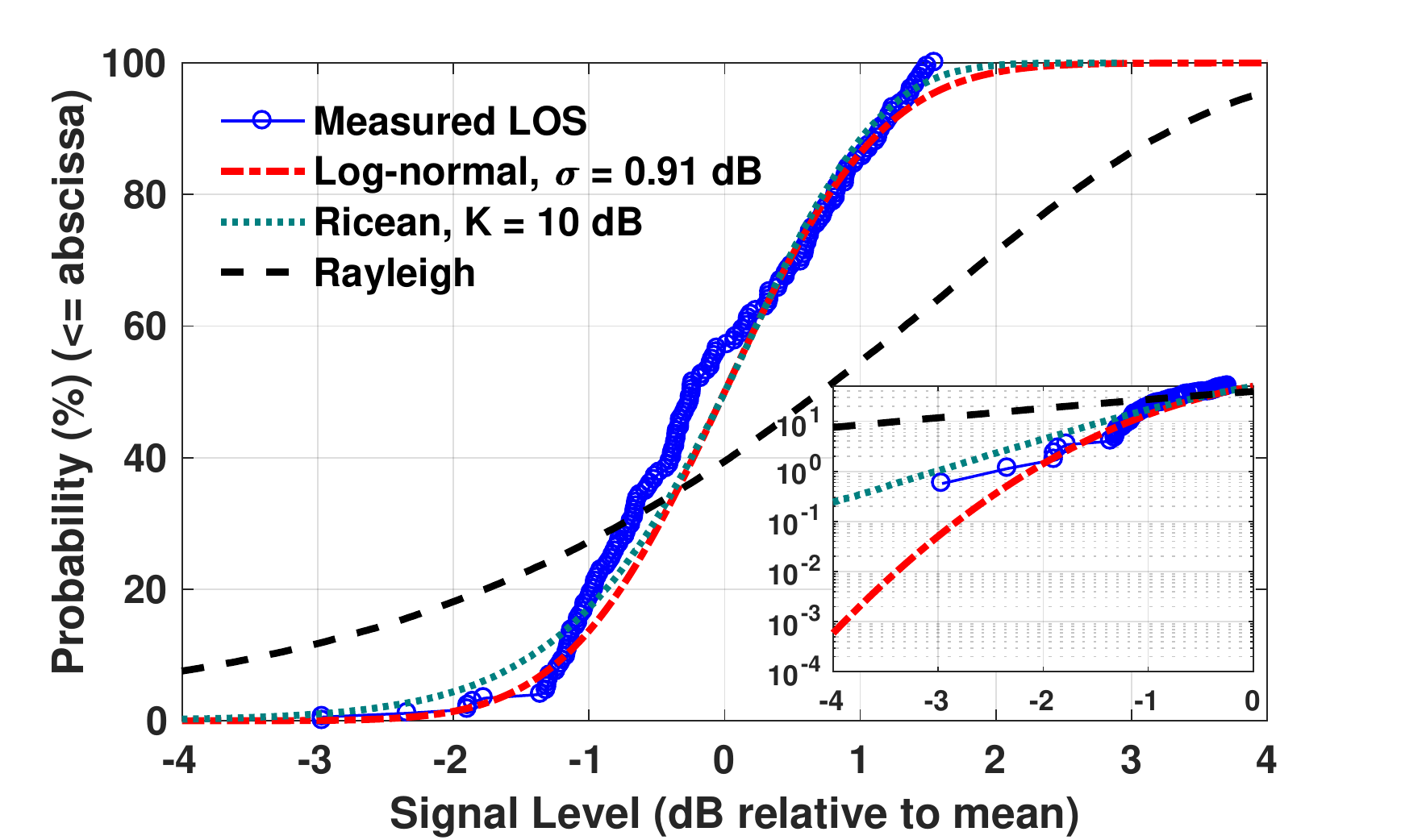}
	\caption{CDF of the measured small-scale spatial fading distribution of the received voltage amplitude for the omnidirectional RX antenna pattern in the LOS environment at 73 GHz with a 1 GHz RF bandwidth.}
	\label{fig:omniLOSFading}
	\squeezeup
\end{figure}

\begin{figure}
	\centering
	\includegraphics[width=2.7in]{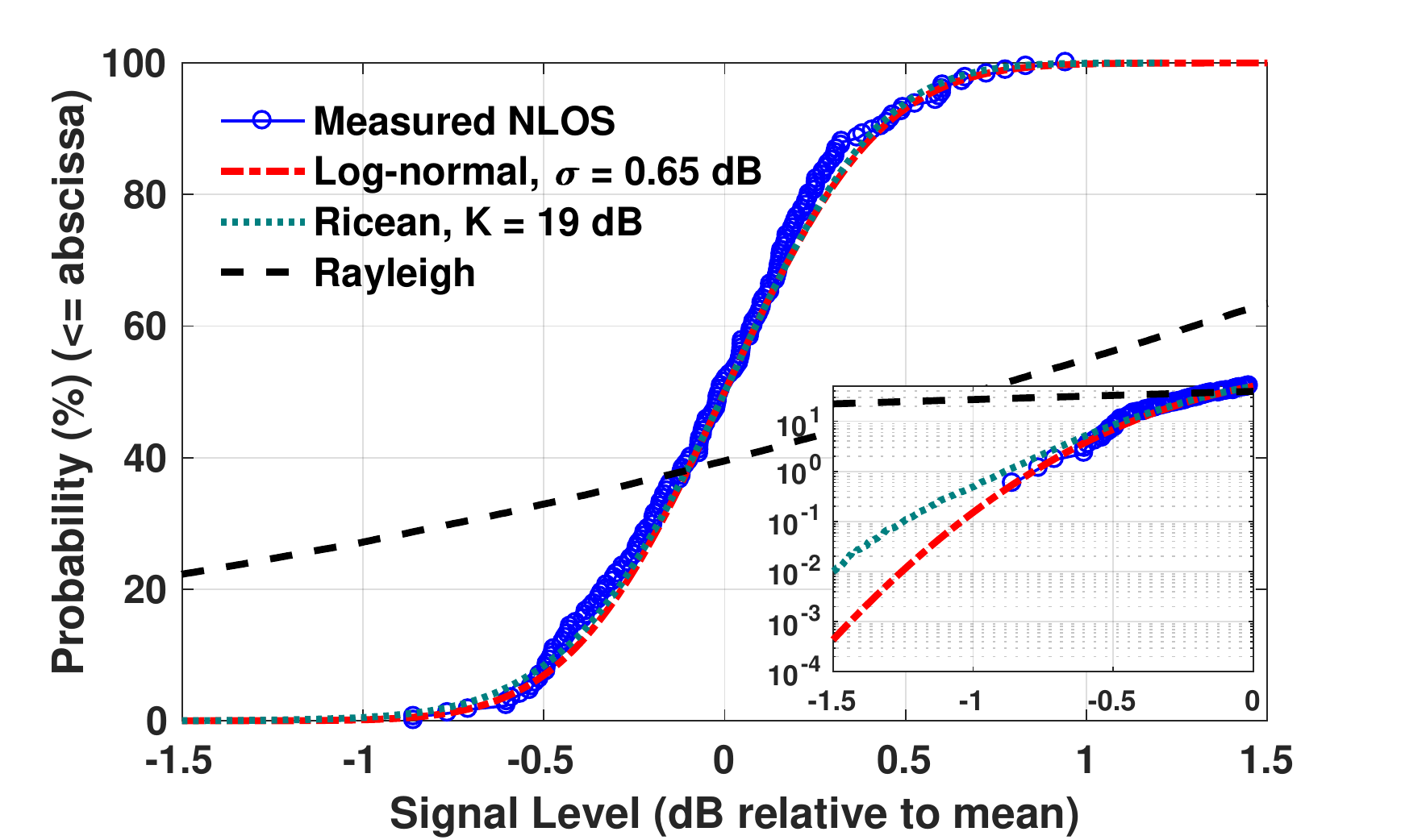}
	\caption{CDF of the measured small-scale spatial fading distribution of the received voltage amplitude for the omnidirectional RX antenna pattern in the NLOS environment at 73 GHz with a 1 GHz RF bandwidth.}
	\label{fig:omniNLOSFading}
	\squeezeup
\end{figure}

Fig.~\ref{fig:omniLOSFading} illustrates the cumulative distribution function (CDF) of the measured small-scale received voltage amplitude at 73 GHz with a 1 GHz RF bandwidth over the 35.31-cm length track with 175 track positions in increments of half-wavelength (2.04 mm) for the omnidirectional RX antenna pattern in the LOS environment\textcolor{black}{\cite{Sun17a}}. \textcolor{black}{Superimposed with the measured curve are the CDFs of the Rayleigh distribution, the zero-mean log-normal distribution with a standard deviation of 0.91 dB (obtained from the measured data), and the Ricean distribution with a $K$-factor of 10 dB obtained from the measured data by dividing the total received power contained in the LOS path by the power contributed from all the other reflected or scattered paths. As shown in Fig.~\ref{fig:omniLOSFading}, the measured 73 GHz small-scale spatial fading in the LOS environment can be approximated by the Ricean distribution with a $K$-factor of 10 dB, indicating that there is a dominant path (i.e., the LOS path) contributing to the total received power, and that the received signal voltage amplitude varies little over the 35.31-cm (about 87 wavelengths) length track. The log-normal distribution does not fit the measured data well in the regions of -3 to -2.5 dB and +1.2 to +1.5 dB about the mean.} The maximum fluctuation of the received voltage amplitude is merely 3 dB relative to the mean value, whereas the fades are much deeper for the Rayleigh distribution. The physical reason for this is the presence of a dominant LOS path. 
The small-scale spatial fading in the NLOS environment for the omnidirectional RX antenna pattern is illustrated in Fig.~\ref{fig:omniNLOSFading}, \textcolor{black}{and the zero-mean log-normal distribution with a standard deviation of 0.65 dB (obtained from the measured data) is selected to fit the measured result, and Ricean and Rayleigh distributions are also given as a reference\textcolor{black}{\cite{Sun17a}}. As evident from Fig.~\ref{fig:omniNLOSFading}, the measured NLOS small-scale spatial fading distribution matches the log-normal fitted curve almost perfectly. In contrast, the Ricean distribution with $K$ = 19 dB does not fit the measured data as well as the log-normal distribution in the tail region around -0.6 dB to -0.8 dB of the relative mean signal level }\textcolor{black}{(as shown by the inset in Fig.~\ref{fig:omniNLOSFading}), since the Ricean $K$ = 19 dB distribution predicts more occurrences of deeper fading events, whereas the log-normal distribution with a 0.65 dB standard deviation predicts a more compressed fading range of -0.8 dB to +0.8 dB about the mean, which was observed for the wideband NLOS signals.} The fact that the local fading of received voltage amplitudes in the NLOS environment is \textcolor{black}{log-normal} instead of Rayleigh is similar to models in~\cite{Rap91b} for urban mobile radio channels. For a NLOS environment, there may not be a dominant path, yet the transmitted broadband signal experiences frequency-selective fading (when the signal bandwidth is larger than the coherence bandwidth of the channel~\cite{Rap02a}). Different frequency components of the signal experience uncorrelated fading, thus it is highly unlikely that all parts of the signal will simultaneously experience a deep fade, and the fades over frequency tend to be very sharp, taking up a small portion of the total power received over the entire signal bandwidth~\cite{Rap15a}. Consequently, the total received power changes very little over a small-scale local area. This is a distinguishing feature of wideband mobile signals as compared to narrowband signals.

Apart from small-scale spatial fading, small-scale spatial autocorrelation is also important for wireless modem design. Spatial autocorrelation characterizes how the received voltage amplitudes correlate at different linear track positions within a local area~\cite{Samimi16b}. Spatial autocorrelation coefficient functions can be calculated using Eq.~\eqref{sac}, where $X_k$ denotes the $k^{th}$ linear track position, $E[~]$ is the expectation operator where the average of voltage amplitudes is taken over all the positions $X_k$, and $\Delta X$ represents the spacing between different antenna positions on the track\textcolor{black}{\cite{Sun17a}}. 
\begin{figure*}
	\begin{equation}\label{sac}
	\footnotesize
	\rho=\frac{E\big[\big(A_k(X_k)-\overline{A_k(X_k)}\big)\big(A_k(X_k+\Delta X)-\overline{A_k(X_k+\Delta X)}\big)\big]}{\sqrt{E\big[\big(A_k(X_k)-\overline{A_k(X_k)}\big)^2\big]E\big[\big(A_k(X_k+\Delta X)-\overline{A_k(X_k+\Delta X)}\big)^2\big]}}
	\squeezeup
	\end{equation}
	\squeezeup
\end{figure*}

\begin{figure}
	\centering
	\includegraphics[width=2.5in]{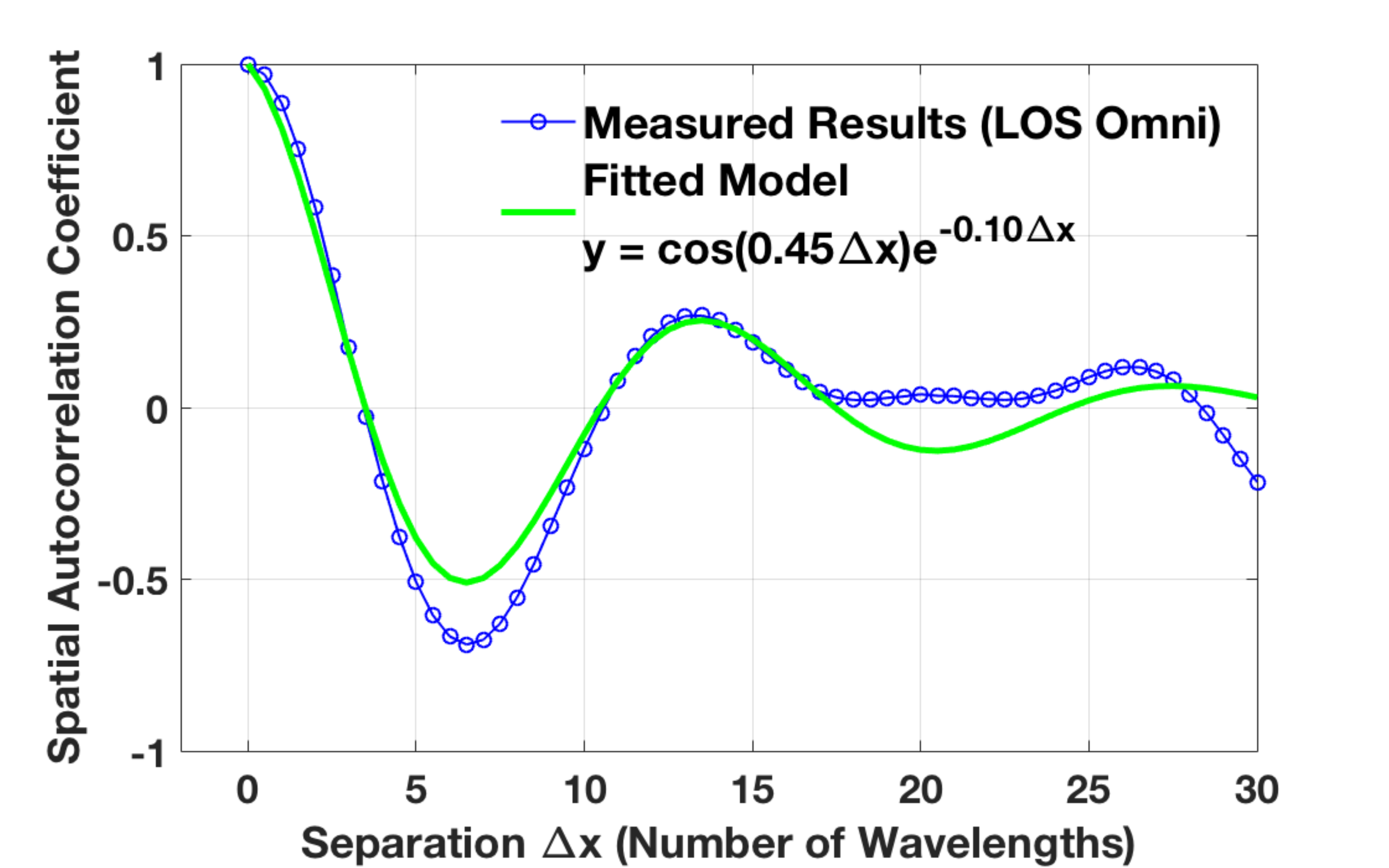}
	\caption{\textcolor{black}{Measured 73 GHz broadband spatial autocorrelation coefficients of the received voltage amplitude in the LOS environment, and the corresponding fitting model. The T-R separation distance is 79.9 m.}}
	\label{fig:omniLOSCor}
	\squeezeup
\end{figure}

\begin{figure}
	\centering
	\includegraphics[width=2.5in]{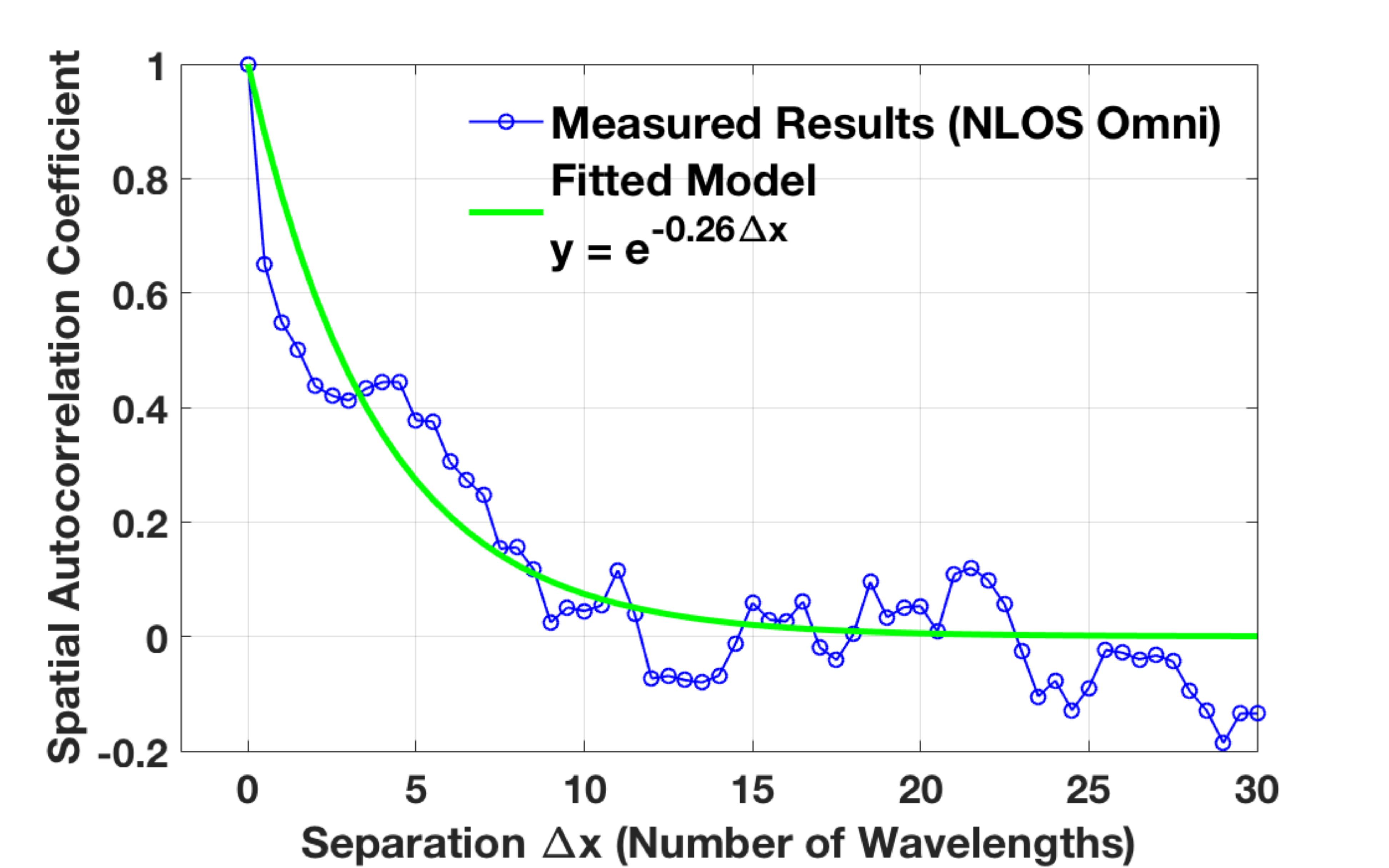}
	\caption{\textcolor{black}{Measured 73 GHz broadband spatial autocorrelation coefficients of the received voltage amplitude in the NLOS environment, and the corresponding fitting model. The T-R separation distance is 75.0 m.}}
	\label{fig:omniNLOSCor}
	\squeezeup
\end{figure}

The measured 73 GHz spatial autocorrelation of the received voltage amplitudes in LOS and NLOS environments with a 1 GHz RF bandwidth are depicted in Fig.~\ref{fig:omniLOSCor} and Fig.~\ref{fig:omniNLOSCor}, respectively. Note that a total of 175 linear track positions over the 35.31-cm length track were measured during the measurements, yielding a maximum spatial separation of 174 half-wavelengths on a single track. Only up to 60 half-wavelengths, however, are shown herein because little change is found thereafter and it provides 100 autocorrelation data points for all spatial separations on a single track, thus improving the reliability of the statistics. According to Fig.~\ref{fig:omniLOSCor}, the received omnidirectional signal voltage amplitude first becomes uncorrelated at a spatial separation of about 3.5$\lambda$, then becomes slightly anticorrelated for separations of 3.5$\lambda$ to 10$\lambda$, and becomes slightly correlated for separations between 10$\lambda$ and 18$\lambda$, and decays towards 0 sinusoidally after 18$\lambda$. Therefore, the spatial correlation can be modeled by a ``damped oscillation'' function of~\eqref{expFit}\textcolor{black}{\cite{Sun17a}}~\cite{Zhang08a}:
\begin{equation}\label{expFit}
\footnotesize
f(\Delta X) = \cos(a\Delta X)e^{-b\Delta X}
\end{equation}
where $\Delta X$ denotes the space between antenna positions, $a$ is an oscillation distance with units of radians/$\lambda$ (wavelength), \textcolor{black}{$T=2\pi/a$} can be defined as the spatial oscillation period with units of $\lambda$ or cm, and $b$ is a constant with units of $\lambda^{-1}$ whose inverse \textcolor{black}{$d=1/b$} is the spatial decay constant with units of $\lambda$. $a$ and $b$ are obtained using the minimum mean square error (MMSE) method to find the best fit between the empirical spatial autocorrelation curve and theoretical exponential model given by~\eqref{expFit}. The ``damped oscillation'' pattern can be explained by superposition of multipath components with different phases at different linear track positions. \textcolor{black}{As the separation distance of linear track positions increases, the phase differences among individual multipath components will oscillate as the separation distance of track positions increases due to alternating constructive and destructive combining of the multipath phases.} This ``damped oscillation'' pattern is obvious in LOS environment where phase difference among individual multipath component is not affected by shadowing effects that occurred in NLOS environments. The form of~\eqref{expFit} also guarantees that the spatial autocorrelation coefficient is always 1 for $\Delta X = 0$, and converges to 0 when $\Delta X$ approximates infinity. The spatial autocorrelation curve for NLOS environment in Fig.~\ref{fig:omniNLOSCor} exhibits a different trend from that in Fig.~\ref{fig:omniLOSCor}, which is more akin to an exponential distribution without damping, but can still be fitted using Eq.~\eqref{expFit} with $a$ set to 0\textcolor{black}{\cite{Sun17a}}. The constants $a$, $b$, are provided in Table~\ref{tbl:ModelPara}, where $T$ is the oscillation period, and $d$ represents the spatial decay constant. From Fig.~\ref{fig:omniNLOSCor} and Table~\ref{tbl:ModelPara} it is clear that after 1.57 cm (3.85 wavelengths at 73.5 GHz) in the NLOS environment, the received voltage amplitudes become uncorrelated (the correlation coefficient decreases to 1/e~\cite{Samimi16a}). We note that Samimi~\cite{Samimi16b} found individual multipath voltage amplitudes received using a 30$^\circ$ Az./El. HPBW antenna became uncorrelated at physical distances of 0.52 cm (0.48 wavelengths at 28 GHz) and 0.67 cm (0.62 wavelengths at 28 GHz) in LOS and NLOS environments, respectively -- smaller decorrelation distances compared to the present 73 GHz results measured using a 60$^\circ$ Az./El. HPBW antenna. 

\begin{table}[t!]
\renewcommand{\arraystretch}{1.4}
\caption{Spatial correlation model parameters in~\eqref{expFit} for 73 GHz, 1 GHz RF bandwidth ($\lambda$=0.41 cm).}~\label{tbl:ModelPara}
\fontsize{8}{6.5}\selectfont
\scriptsize
\begin{center}\scalebox{0.84}{
		\squeezeup
		\begin{tabular}{|>{\centering\arraybackslash}m{1.4cm}|>{\centering\arraybackslash}m{1.3cm}|>{\centering\arraybackslash}m{2.15cm}|>{\centering\arraybackslash}m{1.25cm}|>{\centering\arraybackslash}m{2.15cm}|>{\centering\arraybackslash}m{0.6cm}|}\hline
			\textbf{Condition} & $\boldsymbol{a}~(rad/\lambda)$ & \textcolor{black}{$\boldsymbol{T=2\pi/a}$}	        & \textbf{b}~($\lambda^{-1}$)	 & \textcolor{black}{$\boldsymbol{d=1/b}$}             \\ \hline \hline
			\textbf{LOS Omnidirectional} & {0.45} & {14.0$\lambda$ (5.71 cm)} & {0.10} & {10.0$\lambda$ (4.08 cm)}\\ \hline
			\textbf{NLOS Omnidirectional} & {0} & {\textcolor{black}{Not used}} & {0.26} & {3.85$\lambda$ (1.57 cm)}\\ \hline
			\textbf{LOS Directional} & {0.33 to 0.50} & {12.6$\lambda$ to 19.0$\lambda$ (5.14 cm to 7.76 cm)} & {0.03 to 0.15} & {6.67$\lambda$ to 33.3$\lambda$ (2.72 cm to 13.6 cm)}\\ \hline
			\textbf{NLOS Directional} & {0} & {\textcolor{black}{Not used}} & {0.04 to 1.49} & {0.67$\lambda$ to 25.0$\lambda$ (0.27 cm to 10.2 cm)}\\ \hline
		\end{tabular}}
	\end{center}
\end{table}
	
\subsubsection{Directional Small-Scale Spatial Statistics}
Since mobile devices will use directional antennas, directional statistics are also of interest. In this subsection, we will investigate small-scale spatial fading and autocorrelation of the received voltage amplitudes associated with directional antennas at the RX. 

\begin{figure}
	\centering
	\includegraphics[width=2.8in]{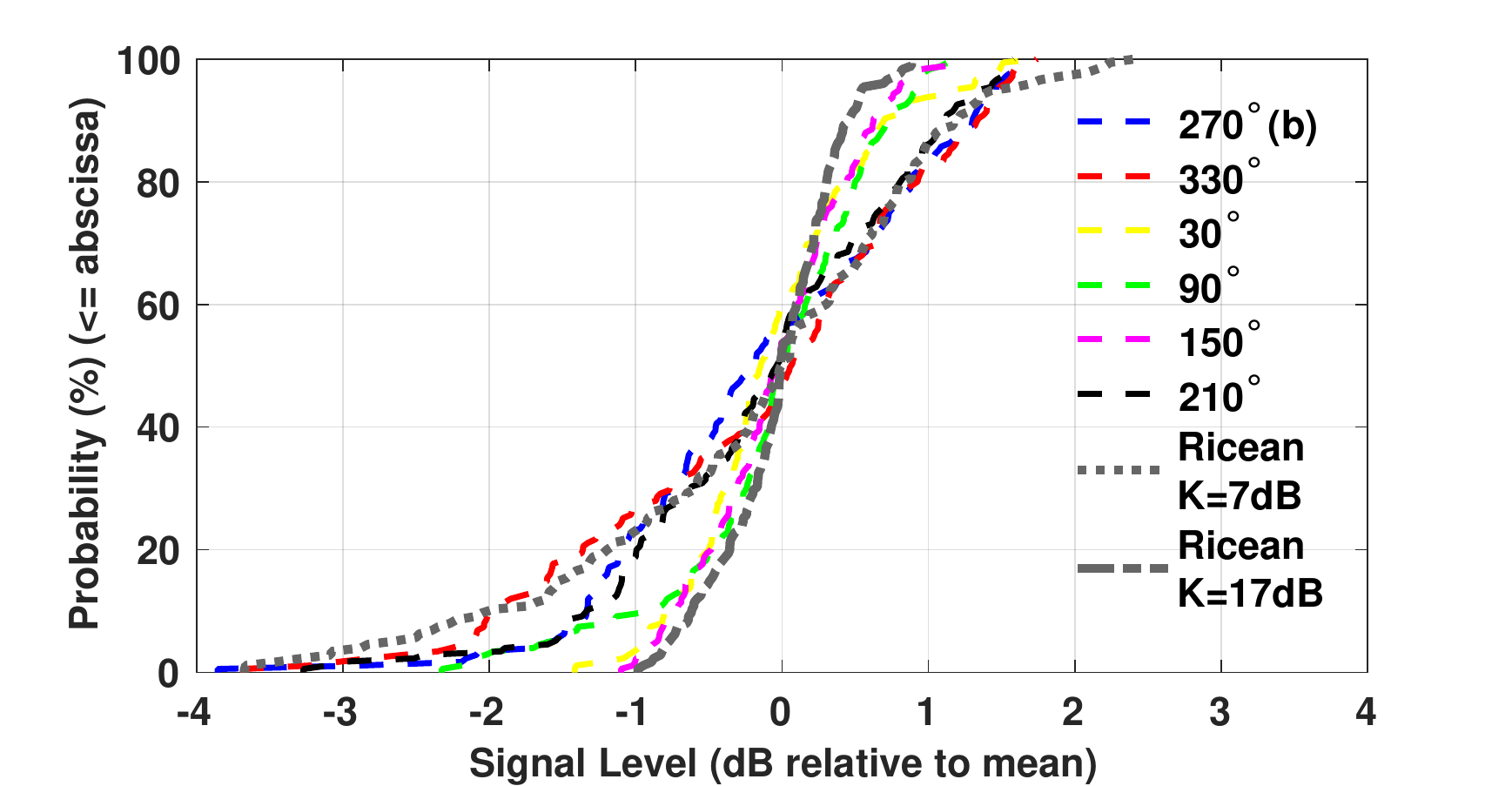}
	\caption{\textcolor{black}{Measured 73 GHz LOS small-scale spatial fading distributions of the directional received voltage amplitude, and the corresponding Ricean fitting curves with different $K$ factors. The angles in the legend denote the receiver antenna azimuth angle, and "b" denotes boresight to the TX.}}
	\label{fig:dirLOSFading}
	\squeezeup
\end{figure}

\begin{figure}
	\centering
	\includegraphics[width=2.8in]{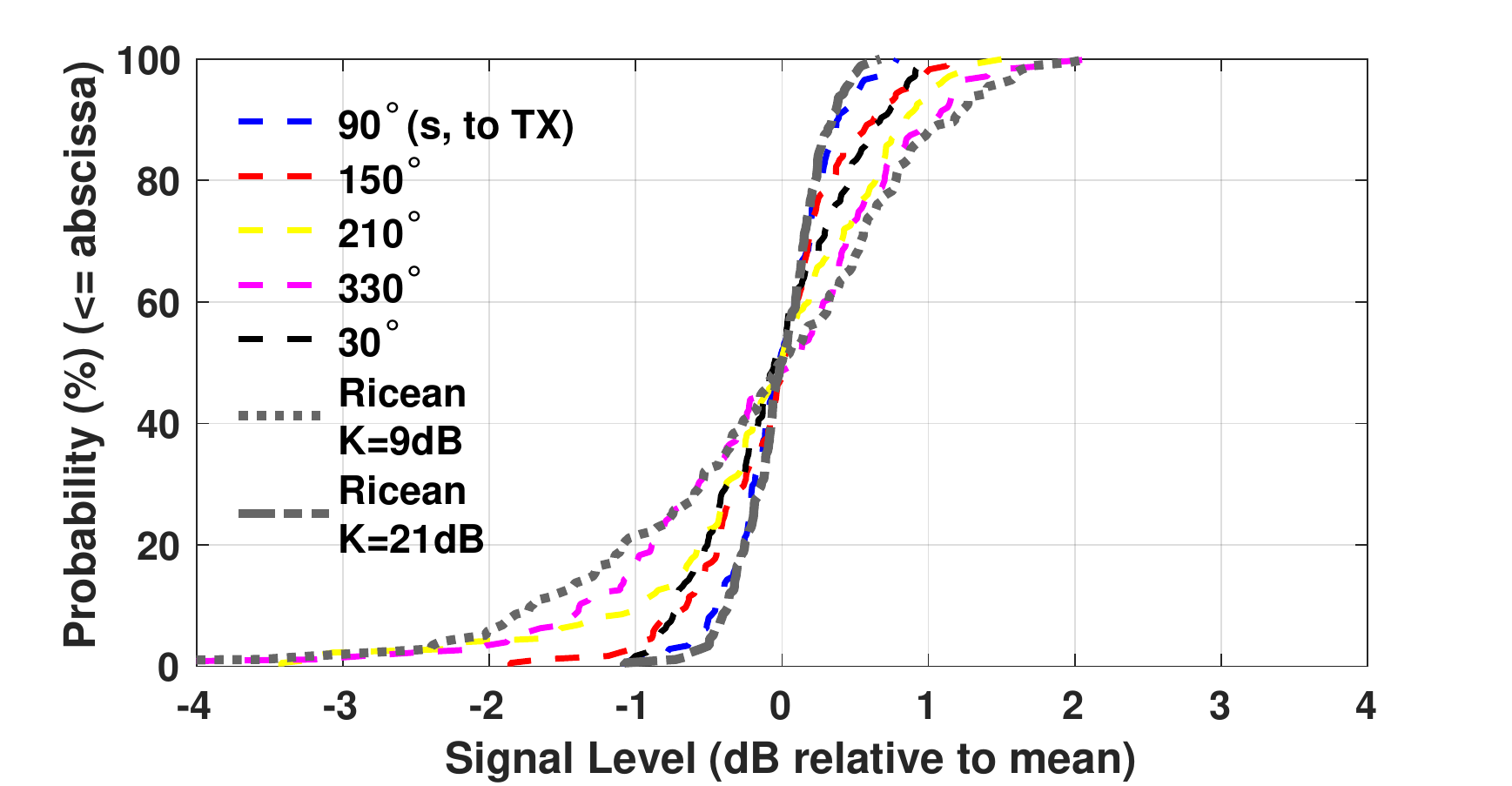}
	\caption{\textcolor{black}{Measured 73 GHz NLOS small-scale spatial fading distributions of the directional received voltage amplitude, and the corresponding Ricean fitting curves with different $K$ factors. The angles in the legend denote the receiver antenna azimuth angle, and "s, to TX" denotes along the direction of the street and pointing to the TX side.}}
	\label{fig:dirNLOSFading}
	\squeezeup
\end{figure}

Small-scale fading of received voltage amplitudes along the linear track using the $7^{\circ}$ Az./El. HPBW TX antenna and $60^{\circ}$ Az./El. HPBW RX antenna in LOS and NLOS environments are shown in Fig.~\ref{fig:dirLOSFading} and Fig.~\ref{fig:dirNLOSFading}, respectively, where the TX and RX were placed as shown in Fig.~\ref{fig:TX_RX_Location}, and each measured curve corresponds to a unique RX antenna azimuth pointing angle relative to true north as specified in the legend. There was no signal for the RX azimuth pointing angle of 270$^{\circ}$ in the NLOS environment, thus the corresponding results are absent in Fig.~\ref{fig:dirNLOSFading}. The strongest pointing directions are 270$^\circ$ and 150$^\circ$ in Figs.~\ref{fig:dirLOSFading} and~\ref{fig:dirNLOSFading}, respectively\textcolor{black}{\cite{Sun17a}}. As shown in Figs.~\ref{fig:dirLOSFading} and~\ref{fig:dirNLOSFading}, the measured directional spatial autocorrelation coefficients resemble Ricean distributions in both LOS and NLOS environments. Possible reason for such distributions is that only one dominant path (accompanied with several weaker paths) is captured by the horn antenna due to its directionality, given the fact that mmWave propagation is directional and the channel is sparse~\cite{Samimi16a}. The Ricean $K$-factor for received voltage amplitudes for various RX pointing directions ranges from 7 dB to 17 dB for the LOS environment, and 9 dB to 21 dB for the NLOS case, as shown in Figs.~\ref{fig:dirLOSFading} and~\ref{fig:dirNLOSFading}, as the RX is moved over a 35.31 cm (86.5 wavelengths at 73.5 GHz) track.

\begin{figure}
	\centering
	\includegraphics[width=3.0in]{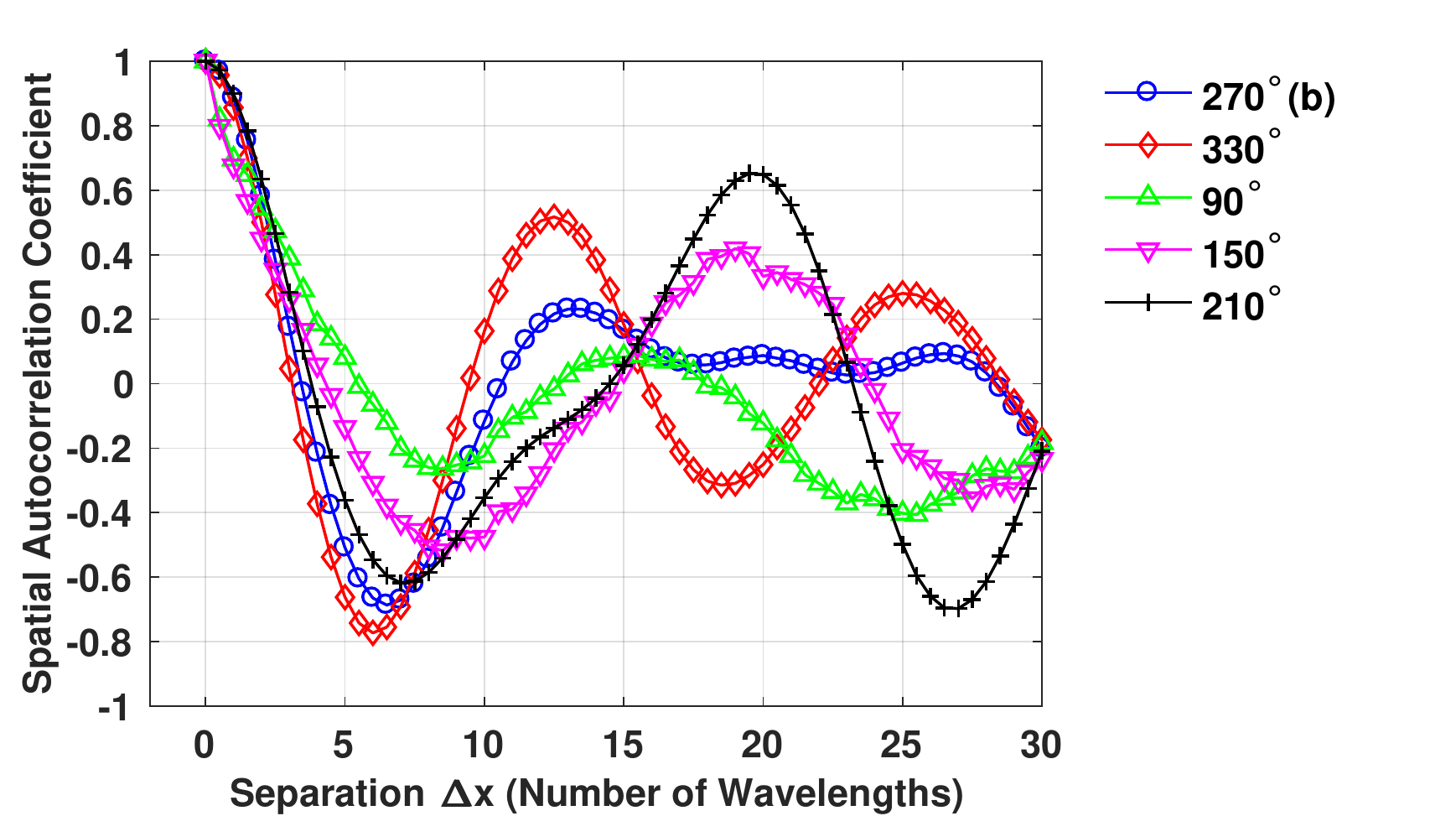}
	\caption{\textcolor{black}{Measured 73 GHz LOS spatial autocorrelation coefficients of the directional received voltage amplitudes. The angles in the legend denote the receiver antenna azimuth angle, and "b" denotes boresight to the TX.}}
	\label{fig:dirLOSCor}
	\squeezeup
\end{figure}

\begin{figure}
	\centering
	\includegraphics[width=3.0in]{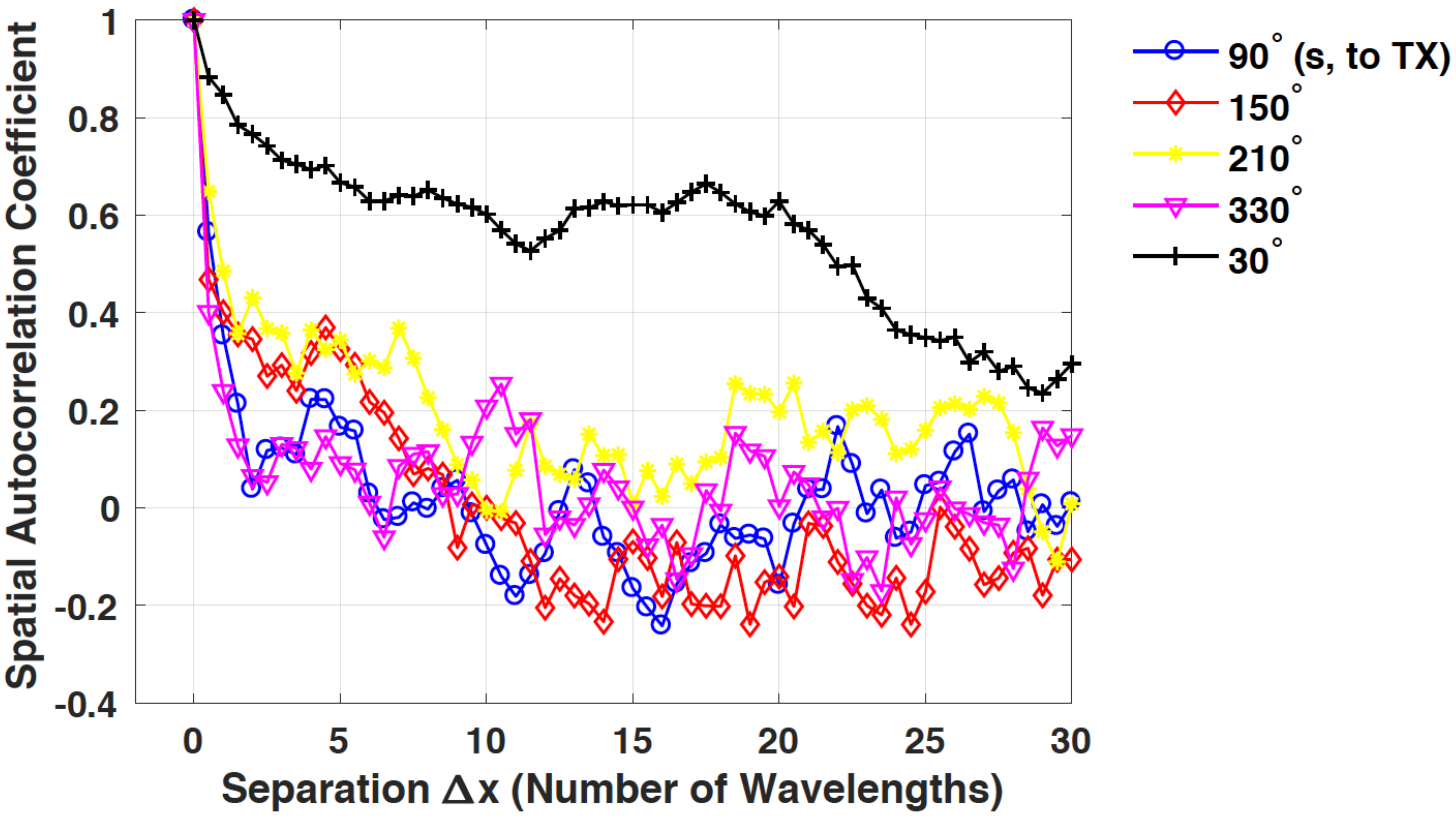}
	\caption{\textcolor{black}{Measured 73 GHz NLOS spatial autocorrelation coefficients of the directional received voltage amplitudes. The angles in the legend denote the receiver antenna azimuth angle, and ``s, to TX" denotes along the direction of the street and pointing to the TX side.}}
	\label{fig:dirNLOSCor}
	\squeezeup
\end{figure}

Figs.~\ref{fig:dirLOSCor} and~\ref{fig:dirNLOSCor} illustrate the spatial autocorrelation coefficients of the received voltage amplitudes for individual antenna pointing angles in LOS and NLOS environments in downtown Brooklyn (see Fig.~\ref{fig:TX_RX_Location}), respectively. As shown by Fig.~\ref{fig:dirLOSCor}, all of the six spatial autocorrelation curves in the LOS environment exhibit sinusoidally exponential decaying trends, albeit with different oscillation patterns and decay rates. The TX-RX boresight-to-boresight pointing angle (270$^\circ$ RX pointing angle, corresponding to the strongest received power) yields the smallest oscillation since there is only a single LOS component in the PDP, while the other pointing directions contain two or more multipath components with varying phases that result in larger oscillation\textcolor{black}{\cite{Sun17a}}. The spatial decay constants for all the curves in Fig.~\ref{fig:dirLOSCor} are given in Table~\ref{tbl:ModelPara}. Compared with the omnidirectional case displayed in Fig.~\ref{fig:omniLOSCor}, it is clear that for the LOS environment, the spatial autocorrelation of both omnidirectional and directional received voltage amplitude obeys similar distribution, namely, the  sinusoidal-exponential function, with similar decorrelation distances. On the other hand, most of the spatial autocorrelation curves for the directional received voltage amplitude shown in Fig.~\ref{fig:dirNLOSCor} are also in line with that given by Fig.~\ref{fig:omniNLOSCor}. One exception for the 73 GHz spatial correlation is found at 30$^\circ$ pointing angle that is the second strongest pointing direction (see Fig. \ref{fig:dirNLOSCor}), where decorrelation was much more gradual and decreases to 1/e at 25.0$\lambda$ (10.2 cm), probably due to the presence of a dominant path with a relatively constant signal level, likely caused by the diffraction of the southeast corner of Rogers Hall in Fig.~\ref{fig:TX_RX_Location}. It is clear from Fig. 4 in~\cite{Rap16a} that correlation distances vary among typical mmWave measurements~\cite{Rumney16b} due to the site-specific nature of propagation. 

\section{Local Area Channel Transition}
\subsection{Introduction of Local Area Channel Transition}
Large-scale channel characteristics, such as the autocorrelation of shadow fading and delay spread over distance at the mobile, inter-site correlation of shadow fading at the mobile for two base stations or a base station for one mobile, and properties of local area channel transition, play an important role in constructing channel models for wireless communication systems~\cite{Jalden07a,3GPP.38.900,3GPP.36.814,3GPP.36.839}. 
While sufficient studies on large-scale channel characteristics have been conducted at sub-6 GHz frequencies~\cite{Jalden07a,Guan15a,Zhang08a,Kolmonen10a}, similar studies have been rarely conducted at mmWave frequencies. Guan \textit{et al.} investigated spatial autocorrelation of shadow fading at 920 MHz, 2400 MHz, and 5705 MHz in curved subway tunnels~\cite{Guan15a}. Results showed that the 802.16J model was a better fit to the measured data than an exponential model, and the mean decorrelation distances were found to be several meters. Another measurement campaign carried out in urban macro-cell (UMa) environments at 2.35 GHz showed that a double exponential model fit well with the autocorrelation coefficients of shadow fading samples extracted from all measured routes, while for individual routes, an exponentially decaying sinusoid model had better fitting performance~\cite{Zhang08a}. On the other hand, an exponential function was adopted in the 3GPP channel model (Releases 9 and 11) to describe the normalized autocorrelation of shadow fading versus distances~\cite{3GPP.36.814,3GPP.36.839}. Moreover, Kolmonen \textit{et al.}~\cite{Kolmonen10a} investigated interlink correlation of eigenvectors of MIMO correlation matrices based on a multi-site measurement campaign at 5.3 GHz using a bandwidth of 100 MHz. Results showed that the first eigenvectors for both x- and y-oriented arrays were highly correlated when two RX locations were largely separated. In the following subsections, local area and channel transition measurements at 73 GHz are described and analyzed.

\subsection{Measurement System, Environment, and Procedure for Local Area Channel Transition}
The measurement system used for local area channel transition measurements was identical to the one used in the previous small-scale fading and autocorrelation measurements described in Section~\ref{sec:smallscale}, except from the antennas~\cite{Mac17a}\textcolor{black}{\cite{Mac17c}}. A 27 dBi gain (7$^\circ$ azimuth and elevation Az./El. HPBW) and 20.0 dBi gain (15$^\circ$ Az./El. HPBW) antenna were used at the TX and RX, respectively. Detailed specifications regarding the measurement system are provided in Table \ref{tbl:2}.

The local area channel transition measurements were conducted at 73 GHz in the MetroTech Commons courtyard next to 2 and 3 MetroTech Center in downtown Brooklyn. During measurements, the TX and RX antennas were set to 4.0 m and 1.5 m above ground level, respectively. For each set of \textit{cluster} or \textit{route} scenario RX locations, the TX antenna remained fixed and pointed towards a manually selected azimuth and elevation pointing angle that resulted in the strongest received power at the starting RX position (RX81 for \textit{route} measurements, and RX51 and RX61 for the LOS and NLOS \textit{cluster} measurements, respectively). For each specific TX-RX combination, five consecutive and identical azimuth sweeps ($\sim$3 minutes per sweep and $\sim$2 minutes between sweeps) were conducted at the RX in HPBW step increments ($15^\circ$) where a PDP was recorded at each RX azimuth pointing angle and resulted in at most 120 PDPs ($\frac{360}{15}\times5=120$) per combination (some angles did not have detectable signal above the noise). The best RX pointing angle in the azimuth plane was selected as the starting point for the RX azimuth sweeps (elevation remained fixed for all RX's), at each RX location measured. 

\begin{figure}
	\begin{center}
		\includegraphics [width = 0.37\textwidth]{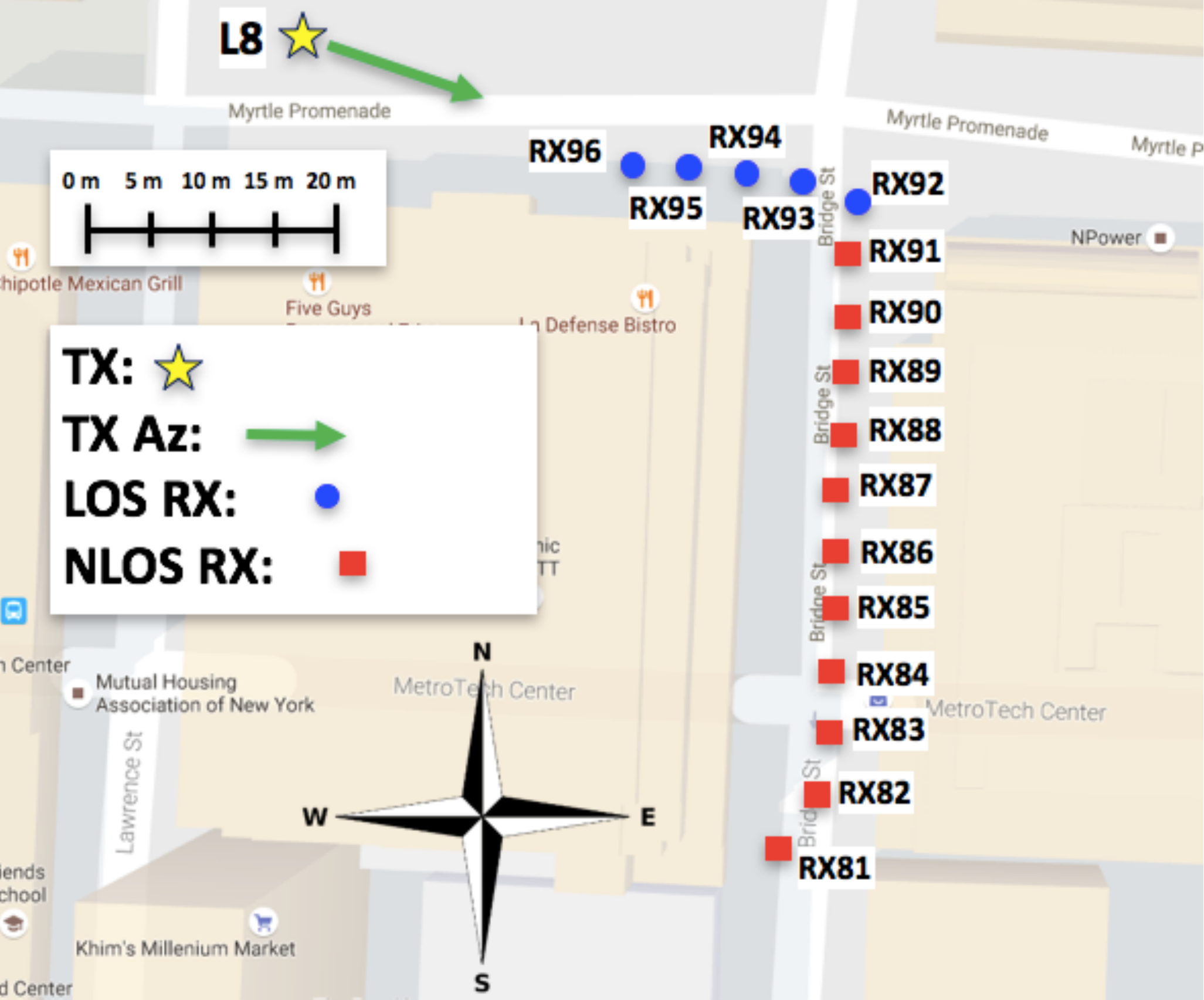}
		\caption{2D map of TX and RX locations for \textit{route} NLOS to LOS transition measurements. The yellow star is the TX location, blue dots represent LOS RX locations, and red squares indicate NLOS RX locations. $\textrm{N}=0^\circ$. }\label{fig:Map_Large_Scale_C1}
	\end{center}
	\squeezeup
\end{figure}

For the \textit{route} measurements, 16 RX locations were measured for a fixed TX location (L8) with the RX locations positioned in 5 m adjacent increments of each other forming a simulated route in the shape of an ``L" around a building corner from a LOS to NLOS region, as provided in Fig.~\ref{fig:Map_Large_Scale_C1}. The LOS location (five: RX92 to RX96) and NLOS location (11: RX81 to RX91) T-R separation distances (Euclidean distance between TX and RX) varied from 29.6 m to 49.1 m and 50.8 m to 81.5 m, respectively. The TX antenna at L8 kept the same azimuth and elevation pointing angle of 100$^\circ$ and 0$^\circ$, respectively, during each experiment (see Fig.~\ref{fig:Map_Large_Scale_C1}. Therefore, the LOS measurements have the TX and RX antennas roughly on boresight in the LOS situation, but they are not exactly on boresight throughout the entire experiment over all measured locations. The general layout of measurements consisted of the RX location starting at RX81, approximately 54 m along an urban canyon (Bridge Street: 18 m width), with the TX antenna pointed towards the opening of the urban canyon (see Fig.~\ref{fig:Map_Large_Scale_C1}). The LOS locations were in clear view of the RX, but with some nearby minor foliage and lamppost obstructions. 

\begin{figure}
	\begin{center}
		\includegraphics [width = 0.37\textwidth]{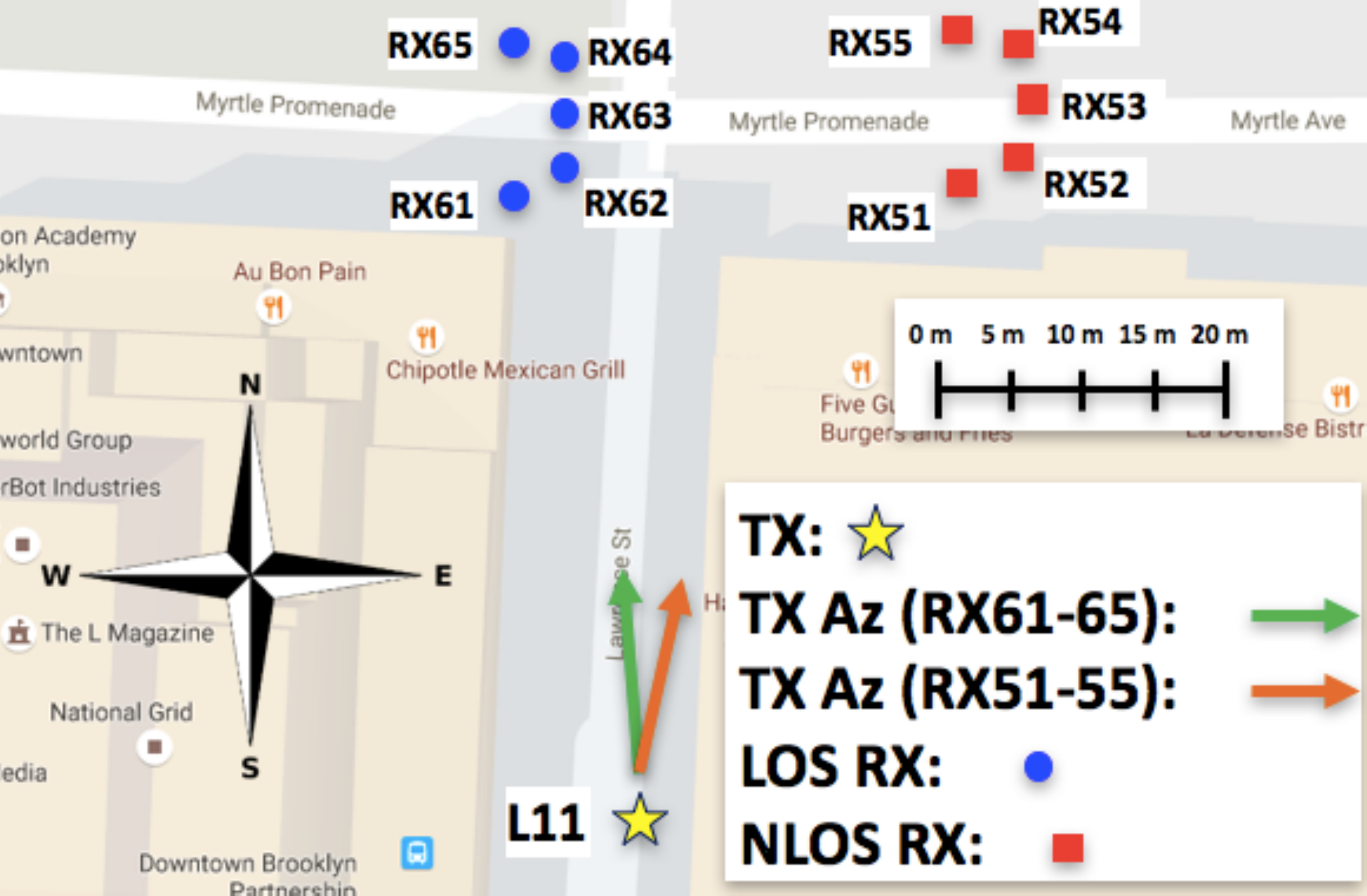}
		\caption{2D map of TX and RX locations for \textit{cluster} measurements with LOS and NLOS RX clusters. The yellow star is the TX location, blue dots represent LOS RX locations, and red squares indicate NLOS RX locations. Pointing to the top is $0^\circ$.}\label{fig:Map_Large_Scale_C2}
		\squeezeup
	\end{center}
	\squeezeup
\end{figure}

For the \textit{cluster} measurements, 10 RX locations were measured for a fixed TX location (L11), with two sets of RX clusters, one in LOS (RX61 to RX65) and the other in NLOS (RX51 to RX 55). For each cluster of RX's, the adjacent distance between each RX location was 5 m, however, the path of adjacent RX locations took the shape of a semi-circle as displayed in Fig.~\ref{fig:Map_Large_Scale_C2}. The LOS cluster T-R separation distances (Euclidean distance between TX and RX) varied between 57.8 m and 70.6 m with a fixed TX antenna azimuth and elevation departure angle of 350$^\circ$ and -2$^\circ$, respectively, and fixed RX elevation angles of $+3^\circ$, to ensure rough elevation and azimuth alignment for all RX locations. For the NLOS cluster, the T-R separation distances were between 61.7 m and 73.7 m with a fixed TX antenna azimuth and elevation departure angle of 5$^\circ$ and -2$^\circ$, respectively, and fixed RX elevation angles of $+3^\circ$. The LOS cluster of RX's was located near the opening of an urban canyon near some light foliage, while the TX location was $\sim$57 m along an urban canyon (Lawrence Street: 18 m width). The NLOS cluster of RX locations was around the corner of the urban canyon opening in a courtyard area (see Fig.~\ref{fig:Map_Large_Scale_C2}), also with nearby moderate foliage and lampposts.

\subsection{Local Area Channel Transition Results and Stationarity}
\begin{figure}[t!]
	\begin{center}
		\includegraphics [width = 0.37\textwidth]{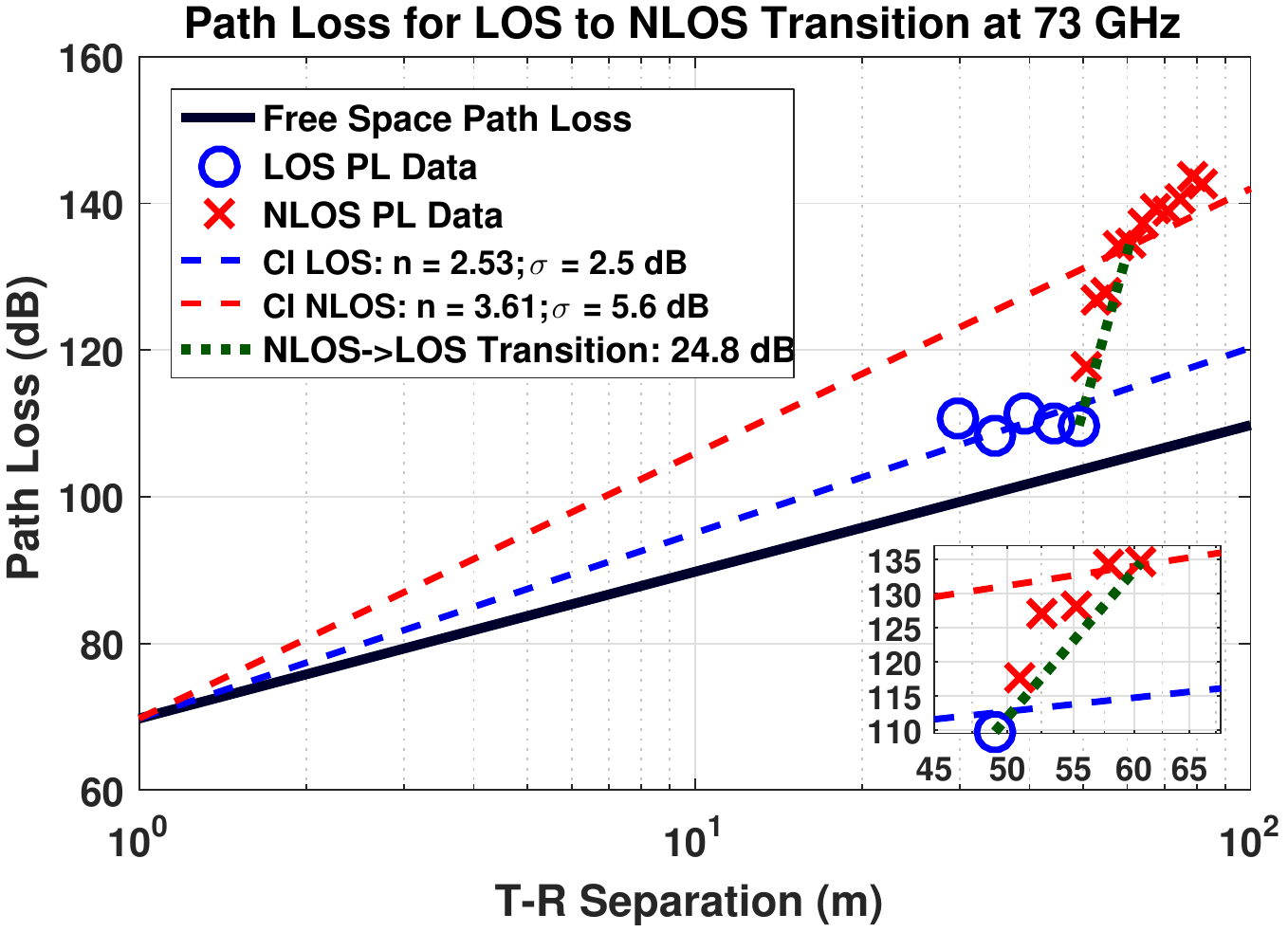}
		\caption{Omnidirectional path loss for route measurements for an RX transitioning from a NLOS to a LOS region.}\label{fig:routePL}
		\squeezeup
	\end{center}
	\squeezeup
\end{figure}

The route measurements mimicked a person moving along an urban canyon from a NLOS to a LOS region, in order to understand the evolution of the channel during the transition. Fig.~\ref{fig:routePL} displays the omnidirectional path loss for each of the RX locations (RX81 to RX96) where the received power from the individual directional measurements at the RX was summed up to determine the entire omnidirectional received power at each measurement location (out of the 5 sweeps, the maximum power at each angle was used, although variation was less than a dB between sweeps)~\cite{Mac14b,Sun15a}.

The transition from LOS to NLOS in Fig.~\ref{fig:routePL} is quite abrupt, where path loss increases by $\sim$8 dB from RX92 to RX91, similar to the abrupt diffraction loss noticed in Section~\ref{sec:diffraction}. Path loss then increases 9 dB further from RX91 to RX90, 1 dB from RX90 to RX89, 6 dB from RX 89 to RX88, and 1 dB from RX88 to RX 87. This observation shows a large initial drop in 8 dB at the LOS to NLOS transition region, but an overall 25 dB drop in signal power when moving from LOS conditions to deeply shadowed NLOS conditions approximately 25 meters farther along a perpendicular urban canyon ($\sim$10 m increase in Euclidean T-R separation distance), when using an omnidirectional RX antenna. The 25 dB drop in signal strength over a 25 m path around a corner (1 dB/m) is important for handoff considerations. The signal fading rate is 35 dB/s for vehicle speeds of 35 m/s, or 1 dB/s for walking speeds of 1 m/s. This motivates the use of beam scanning and phased array technologies in the handset for urban mobile mmWave communications that will search for and find the strongest signal paths~\cite{Sun14b}, and future work will study the best antenna pointing angles at each location from these measurements.  

\begin{figure}
	\centering
	\begin{subfigure}[b]{0.39\textwidth}
		\centering
		\includegraphics[trim={0 0.5cm 0 0},clip,width=\textwidth]{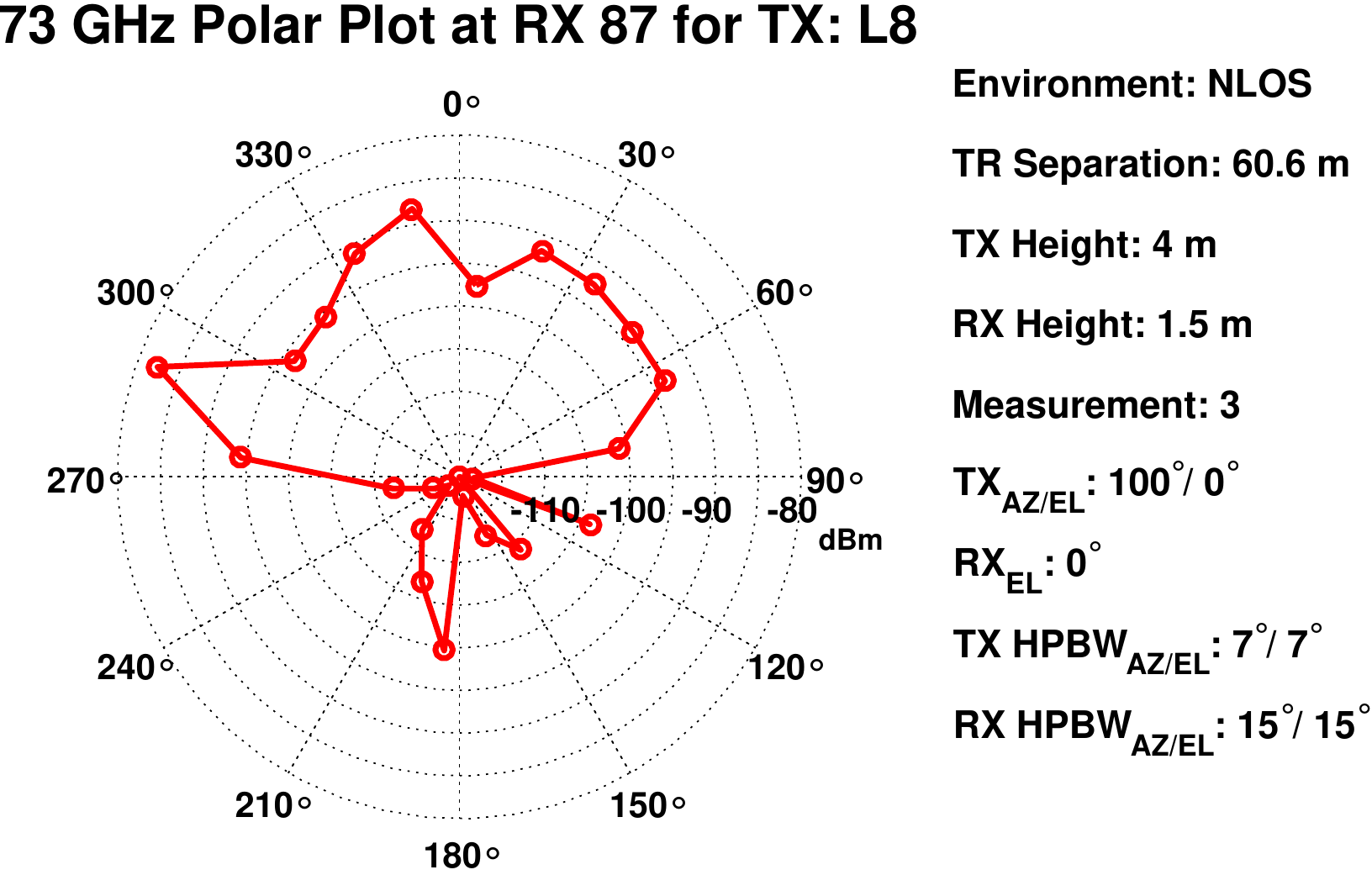}
		\caption[]%
		{{\small RX87: NLOS}}   
		\label{fig:RX87}
	\end{subfigure}
	\begin{subfigure}[b]{0.39\textwidth}   
		\centering 
		\includegraphics[trim={0 0.5cm 0 0},clip,width=\textwidth]{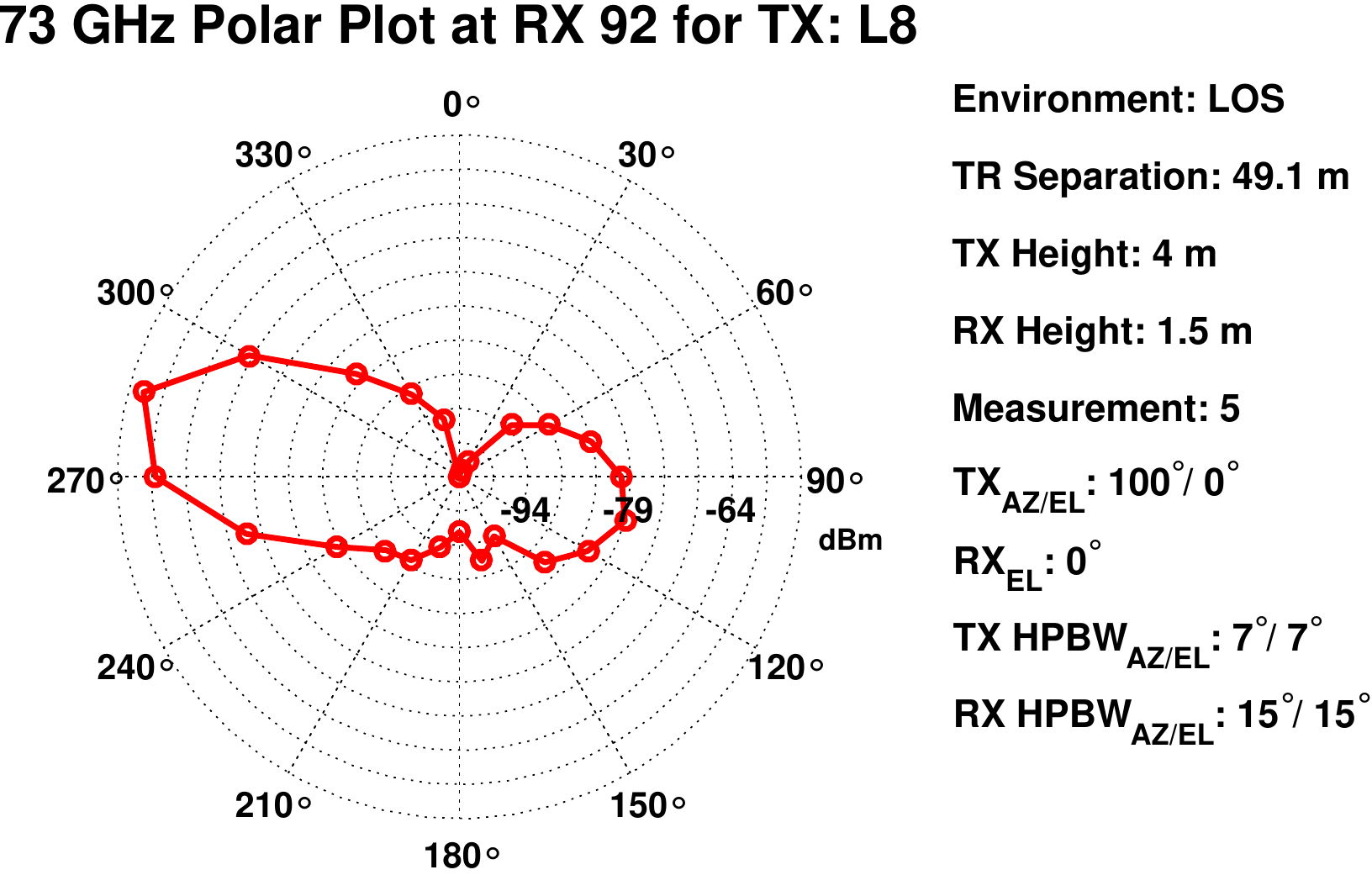}
		\caption[]%
		{{\small RX92: LOS}}    
		\label{fig:RX92}
	\end{subfigure}
	\caption{\small Route scenario polar plots of RX azimuth spectra for RX 87 (NLOS location) and RX 92 (LOS location) that show the evolution of AOA energy around a corner.} 
	\label{fig:case1Polar}
	\squeezeup
\end{figure}

\textcolor{black}{Azimuth power spectra are useful to study how the arriving energy changes as an RX moves from NLOS to LOS. Work in~\cite{Samimi16a} showed that energy arrives in directional lobes in mmWave channels. Fig.~\ref{fig:case1Polar} displays an RX polar plot from RX87 (NLOS location) and RX92 (LOS location), where the TX was pointed in the 100$^\circ$ direction towards the street opening (see Fig.~\ref{fig:Map_Large_Scale_C1}). Fig.~\ref{fig:RX87} shows the power azimuth spectra at RX87 ($\sim$25 meters down the urban street canyon) where energy from the TX waveguides and reflects down Bridge Street such that there is one main broad lobe at the RX oriented in the 0$^\circ$ direction and a small narrow lobe in the 180$^\circ$ direction from weak reflectors and scattering. The large azimuth spread in the main lobe demonstrates the surprisingly reflective nature of the channel at the 73 GHz mmWave band~\cite{Samimi13a,Samimi16a}.}

\textcolor{black}{Fig.~\ref{fig:RX92} displays the power azimuth spectra at RX92 in LOS with a strong central lobe coming from the direction of the TX (285$^\circ$). A relatively strong secondary lobe (100$^\circ$) is also apparent in Fig.~\ref{fig:RX92} with energy contributions from reflections off of the building to the east of RX92 and additional reflectors and scatterers from nearby lampposts and signs.} Table~\ref{tbl:clusterSTD} displays the standard deviations of the omnidirectional received power values measured along the LOS and NLOS routes shown in Fig.~\ref{fig:Map_Large_Scale_C1}. The omnidirectional received power standard deviation of 1.2 dB is generally small for the LOS locations but is much larger in NLOS (7.9 dB) due to substantial scattering along the route of RX locations along the urban canyon where path loss tends to increase non-linearly over log-distance.

\begin{table}
	\centering
	\caption{Omnidirectional received power standard deviation for the large-scale route and cluster scenario measurements.}
	\label{tbl:clusterSTD}
	\begin{center}
		\scalebox{0.78}{
			\begin{tabu}{|c|[1.6pt]c|}
				\hline 
				\textbf{Measurement Set} & \textbf{Omnidirectional Received Power \bm{$\sigma$} [dB]} \\ \specialrule{1.5pt}{0pt}{0pt}
				\textit{Route} - LOS: RX92 to RX96 &	    1.2	\\ \hline
				\textit{Route} - NLOS: RX81 to RX91 &	    7.9	\\ \hline
				\textit{Cluster} - LOS: RX61 to RX65 &	    4.3	\\ \hline
				\textit{Cluster} - NLOS: RX51 to RX55 &	    2.2	\\ \hline
		\end{tabu}}
	\end{center}
	\squeezeup
\end{table}

The \textit{cluster} measurements for the TX at L11 were designed to understand the stationarity of received power in a local area (larger than small-scale distances) on the order of many hundreds to thousands of wavelengths (5 to 10 meters) at mmWave. In LOS, the cluster of five RX locations with a fixed directional TX antenna resulted in an omnidirectional received power standard deviation of 4.3 dB over local area of 5 m x 10 m, a relatively small variation, indicating a reasonably stationary average received power for a local set of RX locations in LOS at 73 GHz. The NLOS cluster resulted in an even lower omnidirectional received power standard deviation of 2.2 dB, over a 5 m x 10 m local area. The small fluctuation in received power over the local area of the LOS and NLOS clusters implies that received power does not significantly vary over RX locations separated by even a few to several meters in a dense urban environment at mmWave. As an aside, the directional CI model path loss exponent (PLE) using a 1 m free space reference in the \textit{route} measurements was 2.53 in LOS (a bit higher than free space due to elevation mismatch over the route) and 3.61 in NLOS (for a single TX beam)~\cite{Mac17a}. Recent work in~\cite{Wang17a} at 28 GHz studied the stationarity of wideband mmWave channels, and reported smaller stationary regions than at 2 GHz. 

\section{Conclusion and Discussion}~\label{sec:conc}
Measurements and analysis were presented on diffraction, human blockage, small-scale fading and local area channel transition at mmWave frequencies. We showed the KED model is suitable for modeling diffraction at cmWave and mmWave bands in indoor scenarios, while a creeping wave linear model is applicable to outdoor environments. \textcolor{black}{Indoor and outdoor diffraction measurements confirm theoretical simulations~\cite{Deng16a} that mmWave diffraction will not be a dominant propagation mechanism. Measurements show significant attenuation of 30 dB in indoor channels and 40 dB or more in outdoor channels in the deep shadow region. Therefore, systems will need to be designed to overcome extremely large fades when a signal path is blocked and when diffraction is no longer viable. MmWave antenna systems at the TX and RX will need to cooperatively search for secondary paths at different pointing directions. The large fades due to diffraction loss have implications on the design of physical layer protocols and frame structures to maintain a link, while finding other spatial paths.}

Human blockage measurements showed a person can induce more than 40 dB of loss when standing 0.5 m from the TX or RX antenna, with a signal decay rate of 0.4 dB/ms at walking speeds. The DKED-AG model given here incorporates directional antenna patterns to accurately predict the upper and lower envelopes of measured received power during a blockage \textcolor{black}{which better agrees with real-world measurements when compared to the 3GPP/METIS blockage model (that was shown to underestimate human blockage, most severely when close to the TX or RX)}. \textcolor{black}{Scintillation with $\sim$2 dB peak-to-peak amplitudes} can be seen in the measurements and model just before the blocker enters the field of view of the RX, suggesting that deep fades may be predictable just before they occur. This information may be used in the design of beam steering or handoff algorithms.

\textcolor{black}{Small-scale spatial fading statistics of received signal voltage amplitudes of 1 GHz bandwidth signals at 73 GHz were measured over a 35.31-cm ($\sim$ 87 wavelengths) linear track and show only small power variations when such a wide bandwidth is used. Fading in LOS locations for omnidirectional RX antennas obeyed the Ricean distribution with a $K$-factor of 10 dB, while fading in NLOS locations can be described by the log-normal distribution with a standard deviation of 0.65 dB. The fading depth ranges from -3 dB to +1.5 dB relative to the mean for LOS, and -0.8 dB to +0.8 dB for NLOS. For a 60$^\circ$ directional RX antenna, fading in both LOS and NLOS environments follows the Ricean distribution, where the $K$-factor ranges from 7 dB to 17 dB for LOS, and 9 dB to 21 dB for NLOS, depending on the RX orientation in relation to the environment and the TX~\cite{Sun17a}, while the fading depth varies between -4 dB to +2 dB relative to the mean for both LOS and NLOS environments.}

\textcolor{black}{Spatial autocorrelation modeling of instantaneous total received signal voltage amplitudes showed that the sinusoidal-exponential distribution fits measurements in the LOS environment for both omnidirectional and directional RX antennas. In the NLOS environment, the spatial autocorrelation can be modeled by the exponential distribution for both omnidirectional and directional RX antennas.} Table~\ref{tbl:ModelPara} shows the oscillation distance/period and spatial decay constant to represent the autocorrelation, where rapid decorrelation of received voltage amplitudes occurred over 0.67 to 33.3 wavelengths (0.27 cm to 13.6 cm), depending on the RX orientation in relation to the environment and the TX~\cite{Sun17a}. \textcolor{black}{The short correlation distance, in general, is favorable for spatial multiplexing in MIMO since it allows for uncorrelated spatial data streams to be transmitted from closely-spaced (a fraction to several wavelengths) antennas~\cite{Sun14b}. Furthermore, the small fading depth indicates that the signal quality of an established link between a TX and an RX will not vary much when the RX moves within a local area on the order of a few tens of wavelengths (within 13.6 cm), as shown in Figs.~\ref{fig:dirLOSCor} and~\ref{fig:dirNLOSCor}.}

Local area channel transition measurements show an initial 8 dB power loss and an overall 25 dB power loss when an RX moves from LOS to NLOS around a building corner and along a street canyon in an urban microcell environment. However, local area path loss measurements (cluster measurements) suggest omnidirectional received power has a relatively stationary mean over a relatively large area (5 m x 10 m) in LOS and NLOS scenarios, respectively, as indicated in Table~\ref{tbl:clusterSTD}. Wideband mmWave signals in typical indoor and outdoor environments were shown to fade at rates from 0.4 dB/ms to 40 dB/s, depending on the speed and environment. The results presented here will aid the 5G wireless community as it develops models for small-scale fading and spatial consistency for handoff algorithms and beam scanning techniques.

\bibliographystyle{IEEEtran}
\bibliography{MacCartney_Bibv10}
\end{document}